\documentclass[journal,twoside]{IEEEtran}
\usepackage{cite}
\ifCLASSINFOpdf
\else
   \usepackage[dvips]{graphicx}
\fi
\usepackage{amsfonts}
\usepackage{lpic}
\usepackage{psfrag}
\usepackage{epsf} 
\usepackage{dsfont}
\usepackage{textcomp}
\usepackage{stmaryrd}
\usepackage{amssymb}
\usepackage{mathrsfs}
\usepackage{textcomp}
\usepackage{lettrine} 
\usepackage{flushend}
\usepackage{amsbsy}

\usepackage[some,bottom]{background}
\SetBgContents{Copyright \copyright  2013 IEEE. Personal use of this material is permitted. However, permission to use this material for any other purposes must be obtained from the IEEE by sending an email to pubs-permissions@ieee.org.}
\SetBgScale{0.65}
\SetBgOpacity{1}
\SetBgColor{black}
\SetBgVshift{1cm}

\newcommand{\tr}[1]{\mathrm{#1}}
\usepackage[cmex10]{amsmath}
\usepackage{accents}
\usepackage{array}
\usepackage{mdwmath}
\usepackage{mdwtab}
\begin{document}
\title{Calculation of the Performance of Communication Systems from Measured Oscillator Phase Noise}
\author{\hspace{0in}\IEEEauthorblockN{M. Reza Khanzadi, \emph{Student Member, IEEE}, Dan Kuylenstierna, \emph{Member, IEEE},\\ Ashkan Panahi, \emph{Student Member, IEEE}, Thomas Eriksson, and Herbert Zirath, \emph{Fellow, IEEE}}
%\\	\IEEEauthorblockA{
%\textdagger{Department of Microtechnology and Nanoscience, Microwave Electronics Lab.}\\
%\textasteriskcentered{Department of Signals and Systems, Communication Systems Group}
%\\{Chalmers University of Technology, Gothenburg, Sweden}
%\\\textit{\{khanzadi, dan.kuylenstierna, thomase, herbert.zirath\}@chalmers.se}}
\thanks{M. Reza Khanzadi is with the Department of 
Signals and Systems, and also the Department of Microtechnology and Nanoscience, Chalmers University of Technology, 41296 Gothenburg, Sweden. 

Thomas Eriksson and Ashkan Panahi are with the Department of 
Signals and Systems, Chalmers University of Technology, 41296 Gothenburg, Sweden.

Dan Kuylenstierna and Herbert Zirath are with the Department of Microtechnology and Nanoscience, Chalmers University of Technology, 41296 Gothenburg, Sweden.

The material in Sec.~\ref{sec_PhaseNoiseStatistics} of this paper was
presented in part at the IEEE International Frequency Control Symposium, Baltimore, MD, May. 2012.
}
}
\BgThispage
\markboth{IEEE TRANSACTIONS ON CIRCUITS AND SYSTEMS—-I: REGULAR PAPERS, 2013}{M. Reza Khanzadi \MakeLowercase{\textit{et al.}}: Calculation of the Performance of Communication Systems from Measured Oscillator Phase Noise}
\maketitle
%{\let\thefootnote\relax\footnote{M. Reza Khanzadi is with the Communication Systems Group, Department of 
%Signals and Systems, and also Microwave Laboratory, Department of Microtechnology and Nanoscience, Chalmers University of Technology, 41296 Gothenburg, Sweden. Thomas Eriksson is with the Communication Systems Group, Department of 
%Signals and Systems, Chalmers University of Technology, 41296 Gothenburg, Sweden.
%Dan Kuylenstierna and Herbert Zirath are with Microwave Laboratory, Department of Microtechnology and Nanoscience, Chalmers University of Technology, 41296 Gothenburg, Sweden. (email: {khanzadi, dan.kuylenstierna, thomase, herbert.zirath}@chalmers.se.)}}\par

\begin{abstract}
Oscillator phase noise (PN) is one of the major problems that affect the performance of communication systems. In this paper, a direct connection between oscillator measurements, in terms of measured single-side band PN spectrum, and the optimal communication system performance, in terms of the resulting error vector magnitude (EVM) due to PN, is mathematically derived and analyzed. First, a statistical model of the PN, considering the effect of white and colored noise sources, is derived. Then, we utilize this model to derive the modified Bayesian Cram\'{e}r-Rao bound on PN estimation, and use it to find an EVM bound for the system performance. Based on our analysis, it is found that the influence from different noise regions strongly depends on the communication bandwidth, i.e., the symbol rate.  For high symbol rate communication systems, cumulative PN that appears near carrier is of relatively low importance compared to the white PN far from carrier. Our results also show that $\boldsymbol{1/f^3}$ noise is more predictable compared to $\boldsymbol{1/f^2}$ noise and in a fair comparison it affects the performance less.
\end{abstract}

\
\IEEEpeerreviewmaketitle
\section{Introduction}
\lettrine[lines=2,nindent=0pt]{O}SCILLATORS are one of the main building blocks in communication systems. Their role is to create a stable reference signal for frequency and timing synchronizations. Unfortunately, any real oscillator suffers from phase noise (PN) which under certain circumstances may be the factor limiting system performance. 

In the last decades, plenty of research has been conducted on better understanding the effects of PN in communication systems \cite{Harris47,Viterbi1963, Pollet1995, Mengali97, Meyr1997, Tomba98, Armada98, Armada2001, Amblard2003151,Munier2003, Zirath2004,Songping2004,Baum2004, Dauwels2004, Panayirci2005, Colavolpe2005, Munier2008,Pedersen2008, Nissila2009, Bhatti2009, Dohler2011,Krishnan2011_1, Krishnan2012_1,Mehrpouyan2012,durisi2013capacity, Krishnan2013_1, martalo2013, ghozlan2013wiener, ghozlan2013multi,Khanzadi2013_0_COMP}. The fundamental effect of PN is a random rotation of the received signal constellation that may result in detection errors \cite{Meyr1997, Munier2003}. PN also destroys the orthogonality of the subcarriers in orthogonal frequency division multiplexing (OFDM) systems, and degrades the performance by producing intercarrier interference \cite{Pollet1995, Tomba98, Armada2001, Songping2004, Munier2008}. Moreover, the capacity and performance of multiple-input multiple-output (MIMO) systems may be severely degraded due to PN in the local oscillators \cite{Baum2004 ,Pedersen2008,  Krishnan2012_1, Mehrpouyan2012,Khanzadi2013_0_COMP}. Further, performance of systems with high carrier frequencies e.g., E-band (60-80 GHz) is more severely impacted by PN than narrowband systems, mainly due to the poor PN performance of high-frequency oscillators \cite{ Zirath2004, Dohler2011}.

To handle the effects of PN, most communication systems include a \emph{phase tracker}, to track and remove the PN. Performance of PN estimators/trackers is investigated in \cite{Meyr1997, Amblard2003151, Bay2008}. In \cite{ZhenQi2010, Georgiadis2004}, the performance of a PN-affected communication system is computed in terms of error vector magnitude (EVM) and \cite{Colavolpe2005, Nissila2009, Krishnan2011_1,Krishnan2013_1} have considered symbol error probability as the performance criterion to be improved in the presence of PN. However, in the communication society, effects of PN are normally studied using quite simple models, e.g, the Wiener process \cite{Colavolpe2005, Demir2006,Khanzadi2011,Krishnan2011_1, Krishnan2012_1,Krishnan2013_1,martalo2013, ghozlan2013wiener, ghozlan2013multi}. A true Wiener process does not take into account colored (correlated) noise sources \cite{Demir2002} and cannot describe frequency and time-domain properties of PN properly \cite{Khanzadi2011, Khanzadi2012_1, Yousefi2010}. This shows the necessity to employ more realistic PN models in study and design of communication systems.

Finding the ultimate performance of PN-affected communication systems as a function of oscillator PN measurements is highly valuable for designers of communication systems when the goal is to optimize system performance with respect to cost and performance constraints. From the other perspective, a direct relation between PN figures and system performance is of a great value for the oscillator designer in order to design the oscillator so it performs best in its target application.

In order to evaluate the performance of PN-affected communication systems accurately, models that precisely capture the characteristics of non-ideal oscillators are required. PN modeling has been investigated extensively in the circuits and systems community over the past decades \cite{Leeson1966,McNeill1997,Herzel1998,Hajmiri1998,Hajimiri1999,Klimovitch2000_1,Klimovitch2000_2,Demir2000,Demir2002,Demir2006,Liu2004, abidi2006phase, Chorti2006}. The authors in \cite{ Leeson1966, Hajimiri1999, Chorti2006} have developed models for the PN  based on frequency measurements, where the spectrum is divided into a set of regions with white (uncorrelated) and colored (correlated) noise sources. Similar models have been employed in \cite{Liu2004, Demir2006} to derive some statistical properties of PN in time domain.

Among microwave circuit designers, spectral measurements, e.g., single-side band (SSB) PN spectrum is the common figure for characterization of oscillators. Normally SSB PN is plotted versus offset frequency, and the performance is generally benchmarked at specific offset frequencies, e.g., 100~kHz or 1~MHz\cite{ Huei95,Zirath2004, Gunnarsson2005}. In this perspective, oscillators with lower content of colored noise come better out in the comparison, especially when benchmarking for offset frequencies close to the carrier \cite{ Huei95}.

In this paper, we employ a realistic PN model taking into account the effect of white and colored noise sources, and utilize this model to study a typical point to point communication system in the presence of PN. Note that this is different from the majority of the prior studies (e.g., \cite{Colavolpe2005, Demir2006,Khanzadi2011,Krishnan2011_1, Krishnan2012_1,Krishnan2013_1,martalo2013, ghozlan2013wiener, ghozlan2013multi}), where PN is modeled as the Wiener process, which is a correct model for oscillators with only white PN sources. Before using the PN model, it is calibrated to fit SSB PN measurements of real oscillators. After assuring that the model describes statistical properties of measured PN over the communication bandwidth, an EVM bound for the system performance is calculated. This is the first time that a direct connection between oscillator measurements, in terms of measured oscillator spectrum, and the optimal communication system performance, in terms of EVM, is mathematically derived and analyzed. Comparing this bound for different PN spectra gives insight into how real oscillators perform in a communication system as well as guidelines to improve the design of oscillators.

%The organization and contribution of this paper are as follow: In Sec.~\ref{sec_system}, we first introduce our PN model. Thereafter, the system model of the considered communication system is introduced. In Sec.~\ref{Sec_SystemPerformance}, we find the performance of the PN affected communication system in terms of EVM. To do so, we first drive the modified Bayesian Cram\'{e}r-Rao bound (MBCRB) on the mean square error of the PN estimation and identify the required PN statistics for calculation of the bound. In Sec.~\ref{sec_PhaseNoiseStatistics} we derive the closed-from autocorrelation functions of the PN increments which are the required statistics for calculation of the system performance. Sec.~\ref{Sec_NumericalSimulation} is dedicated to the numerical simulations. First, the PN samples generation for a given SSB phase spectrum measurement is discussed. Later, the generated samples are used in a Monte-Carlo simulation to evaluate the accuracy of the proposed EVM bound. Then, we study how the EVM bound is affected by different parts of PN. To materialize our theoretical results, the proposed EVM is computed for actual measurements and observations are analyzed. Finally, Sec.~\ref{Sec_Conclusions} concludes the paper.
The organization and contribution of this paper are as follow: 
\begin{itemize}
\item	In Sec.~\ref{sec_system}, we first introduce our PN model. Thereafter, the system model of the considered communication system is introduced. 
\item	In Sec.~\ref{Sec_SystemPerformance}, we find the performance of the PN affected communication system in terms of EVM. To do so, we first drive the modified Bayesian Cram\'{e}r-Rao bound (MBCRB) on the mean square error of the PN estimation.  Note that this is the first time that such a bound is obtained for estimation of PN with both white and  colored sources. The required PN statistics for calculation of the bound are identified. Finally, the mathematical relation between the MBCRB and EVM is computed. 
\item	In Sec.~\ref{sec_PhaseNoiseStatistics} we derive the closed-from autocorrelation function of the PN increments that is required for calculation of the MBCRB.  In prior studies (e.g., \cite{McNeill1997,Liu2004,Demir2006}) the focus has been on calculation of the variance of PN increments.  However, we show that for calculation of the system performance, the autocorrelation function of the PN increments is the required statistics. The obtained autocorrelation function is valid for free-running oscillators and also the low-order phase-locked loops (PLLs).
\item	Sec.~\ref{Sec_NumericalSimulation} is dedicated to the numerical simulations. First, the PN sample generation for a given SSB phase spectrum measurement is discussed in brief. Later, the generated samples are used in a Monte-Carlo simulation to evaluate the accuracy of the proposed EVM bound in a practical scenario. Then, we study how the EVM bound is affected by different parts of the PN spectrum. To materialize our theoretical results, the proposed EVM is computed for actual measurements and observations are analyzed. Finally, Sec.~\ref{Sec_Conclusions} concludes the paper.
\end{itemize}

\begin{table}[t]
\caption{Notations}
\label{tab:Notations}
\centering
\begin{tabular}{ l | c}
  \hline                        
  scalar variable & $x$ \\
  vector & $\mathbf{x}$ \\
  matrix & $\mathbf{X}$ \\
$(a,b)^{th}$ entry of matrix &$[\cdot]_{a,b}$\\
continuous-time signal & $x(t)$ \\
discrete-time signal & $x[n]$ \\
statistical expectation & $\mathbb{E}[\cdot]$\\
real part of complex values & $\Re(\cdot)$\\
imaginary part of complex values & $\Im(\cdot)$\\
angle of complex values & $\arg(\cdot)$\\
natural logarithm & $\log(\cdot)$\\
conjugate of complex values & $(\cdot)^*$\\
vector or matrix transpose & $(\cdot)^T$\\
probability density function (pdf)& $f(\cdot)$\\
Normal distribution with mean $\mu$ and variance $\sigma^2$& $\mathcal{N}(x;\mu,\sigma^2)$\\
second derivative with respect to vector $\mathbf{x}$ &$\nabla^2_{\mathbf{x}}$\\
\hline  
\end{tabular}
\end{table}
%\subsection{Notations}
%Italic letters $(x)$ are scalar variables, bold letters $(\mathbf{ x})$ are vectors, bold upper case letters $(\mathbf{X})$ are matrices, $([\mathbf{X}]_{a,b})$ denotes the $(a,b)th$ entry of matrix $\mathbf{X}$, $\mathbb{E}[\cdot]$ denotes the statistical expectation, $\Re(\cdot)$, $\Im(\cdot)$, and $\arg(\cdot)$ are real part, imaginary part, and angle of complex values, and $(\cdot)^*$, and $(\cdot)^T$, denote conjugate, and transpose, respectively.
\section{System Model}
\label{sec_system}
In this section, we first introduce our PN model in continuous-time domain. Then we present the system model of the considered communication system. 
\subsection{Phase Noise Model}
\label{ssec_PhaseNoiseModel}
In time domain, the output of a sinusoidal oscillator with normalized amplitude  can be expressed as
\begin{align}
V(t)=\left(1+a(t)\right)\cos\left(2\pi f_0 t + \phi(t)\right),
\end{align}
where $f_0$ is the oscillator's central frequency, $a(t)$ is the amplitude noise and $\phi(t)$ denotes the PN \cite{Hajmiri1998}. The amplitude noise and PN are modeled as two independent random processes. According to \cite{Hajmiri1998,Klimovitch2000_2} the amplitude noise has insignificant effect on the output signal of the oscillator. Thus, hereinafter in this paper, the effect of amplitude noise is neglected and the focus is on the study of the PN process. 

In frequency domain, PN is most often characterized in terms of single-side-band (SSB) PN spectrum \cite{Hajmiri1998,Demir2006}, defined as
\begin{align}
\label{normalized_spectrum}
\mathcal{L}(f)=\frac{P(f_0+f )}{P_\tr{Total}},
\end{align}
where $P(f_0+f)$ is the oscillator power within $1$~Hz bandwidth around offset frequency $f$ from the central frequency $f_0$, and $P_\tr{Total}$ is the total power of the oscillator. For an ideal oscillator where the whole power is concentrated at the central frequency, $\mathcal{L}(f)$ would be a Dirac delta function at $f=0$, while, in reality, PN results in spreading the power over frequencies around $f_0$. It is possible to show that at high frequency offsets, i.e., far from the central frequency, where the amount of PN is small, the power spectral density (PSD) of PN is well approximated with $\mathcal{L}(f)$ found from measurements \cite{Chorti2006, Klimovitch2000_1, Demir2006},
\begin{align}
S_{\phi}(f)\approx \mathcal{L}(f)~\tr{for~large}~f.
\end{align}
The offset frequency range where this approximation is valid depends on the PN performance of the studied oscillator \cite{decker1999choosing}. It can be shown that the final system performance is not sensitive to low frequency events. Thus, for low frequency offsets, we model $S_\phi(f)$ in such a way that it follows the same slope as of higher frequency offsets. 
\begin{figure*}[t]
\centering
\psfrag{fc}{$f_\tr{corner}$}%
\psfrag{fp}{$\gamma$}%
\psfrag{logf}{$\log(f)$}%
\psfrag{loglf}{$\log(\mathcal{L}(f))$}%
\psfrag{logsf}{$\log(S_\phi(f))$}%
\psfrag{Nf2}{{\color[rgb]{0,1,0}$\frac{K_2}{f^2}$}}%
\psfrag{Nf3}{{\color[rgb]{1,0,0}$\frac{K_3}{f^3}$}}%
\psfrag{Nf0g}[][][0.8]{{\color[rgb]{0,0,1}$K_0$}}%
\psfrag{(a)}{(a)}%
\psfrag{(b)}{(b)}%
\includegraphics [height=2.5in]{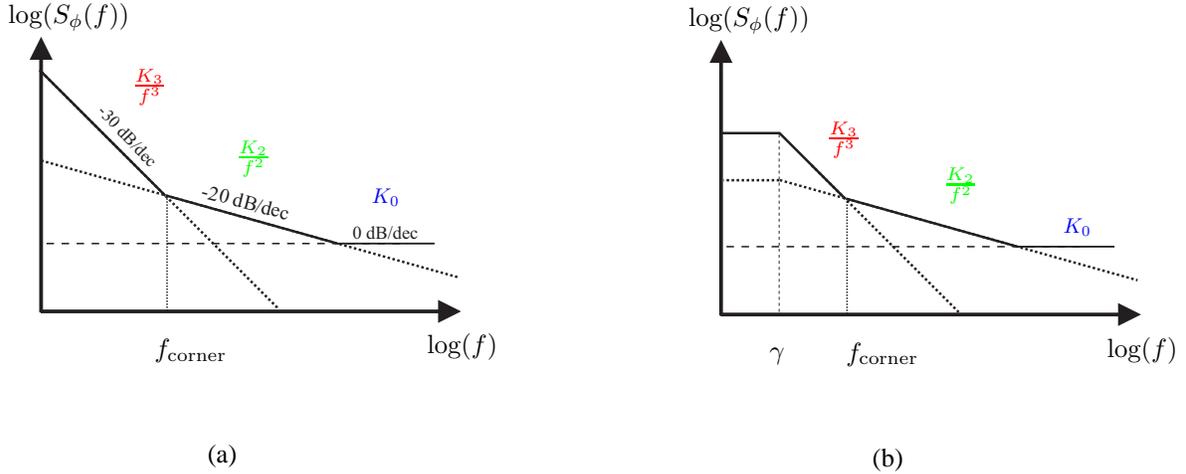}
\caption{Phase noise PSD of a typical oscillator. (a) shows the PSD of a free running oscillator. (b) is a model for the PSD of a locked oscillator, where $\gamma$ is the PLL loop's bandwidth. It is considered that the PN of the reference oscillator is negligible compared to PN of the free running oscillator.}
\label{fig:2}
\end{figure*}
%In experimental data from oscillators, $\mathcal{L}(f)$ normally follows slopes of $-30~\tr{dB}/\tr{decade}$ and $-20~\tr{dB}/\tr{decade}$, until a flat noise floor is reached at higher frequency offsets (see Fig.~\ref{fig:2}-a). The origins of $-30$ and $-20~\tr{dB}/\tr{decade}$ slopes are upconverted flicker noise $(1/f)$ and white noise inside oscillator circuitry \cite{Klimovitch2000_2,Demir2006,Chorti2006}. The flat noise floor, also known as white PN, at higher off-set frequencies, is generally inversely proportional to the signal power of the oscillator \cite{Leeson1966}. For instance, in a measured SSB PN spectrum, the level is set by the noise floor of the measurement system relative to the oscillator's power. In special cases, where a high level of white noise, e.g., shot noise is generated inside the oscillator, it may dominate the system noise floor.
%
In experimental data from free running oscillators, $\mathcal{L}(f)$ normally follows slopes of $-30~\tr{dB}/\tr{decade}$ and $-20~\tr{dB}/\tr{decade}$, until a flat noise floor is reached at higher frequency offsets.

According to Demir's model \cite{Demir2006}, oscillator PN originates from the white and colored noise sources inside the oscillator circuitry.  We follow the same methodology and model PN as a superposition of three independent processes
\begin{align}
\label{PN_parts}
 \phi(t)=\phi_{3}(t)+\phi_{2}(t)+\phi_{0}(t),
\end{align}
where $\phi_{3}(t)$ and $\phi_{2}(t)$  model PN with $-30$ and $-20~\tr{dB}/\tr{decade}$ slopes that originate from integration of flicker noise $(1/f)$ (colored noise) and white noise, denoted as $\Phi_3(t)$ and $\Phi_2(t)$, respectively. Further, $\phi_{0}(t)$ models the flat noise floor, also known as white PN, at higher offset frequencies, that originates from thermal noise and directly results in phase perturbations. In logarithmic scale, the PSD of $\phi_{3}(t)$, $\phi_{2}(t)$, and $\phi_{0}(t)$ can be represented as power-law spectrums \cite{Chorti2006}:
\begin{align}
\label{PN_pawerLaw}
 S_{\phi_3}(f)=\frac {K_3}{f^3},\quad  S_{\phi_2}(f)=\frac {K_2}{f^2}, \quad S_{\phi_0}(f)={K_0},
\end{align}
where $K_3$, $K_2$ and $K_0$  are the PN levels that can be found from the measurements (see Fig.~\ref{fig:2}-a). 

In many practical systems, the free running oscillator is stabilized by means of a phase-locked loop (PLL). A PLL architecture that is widely used in frequency synchronization consists of a free running oscillator, a reference oscillator, a loop filter, phase-frequency detectors and frequency dividers \cite{McNeill_thesis,kundert2003predicting,Demir2006,liu2006jitter}. Any of these components may contribute to the output PN of the PLL. However, PN of the free-running oscillator usually has a dominant effect \cite{McNeill_thesis}. A PLL behaves as a high-pass filter for the free running-oscillator's PN, which attenuates the oscillator's PN below a certain cut-off frequency. As illustrated in Fig~\ref{fig:2}-b, above a certain frequency, PSD of the PLL output is identical to the PN PSD of the free-running oscillator, while below this frequency it approaches a constant value \cite{McNeill_thesis,kundert2003predicting,Demir2006,liu2006jitter}.

\begin{figure}[t]
\centering
\psfrag{OSC}[cc][][1]{Oscillator}%
\psfrag{delay}[cc][][0.6]{Delay~$T$}%
\psfrag{z}[cc][][0.6]{$\zeta_3(t)+\zeta_2(t)$}%
\psfrag{ph0}[cc][][0.6]{$\phi_0(t)$}%
\psfrag{ph2}[cc][][0.6]{$\phi_2(t)$}%
\psfrag{ph3}[cc][][0.6]{$\phi_3(t)$}%
\psfrag{phtp}[cc][][1]{$~$}%
\psfrag{pht}[cc][][1]{$\phi(t)$}%
\psfrag{ej}[cc][][1]{$e^{j(\cdot)}$}%
\psfrag{out}[cc][][1]{Output}%
\psfrag{C}[cc][][0.6]{Circuit}%
\psfrag{N}[cc][][0.6]{Noise}%
\psfrag{H}[cc][][1.5]{$\mathcal{\int}$}%
\psfrag{probe}[cc][][1]{Probe}%
\includegraphics [width=3.4in]{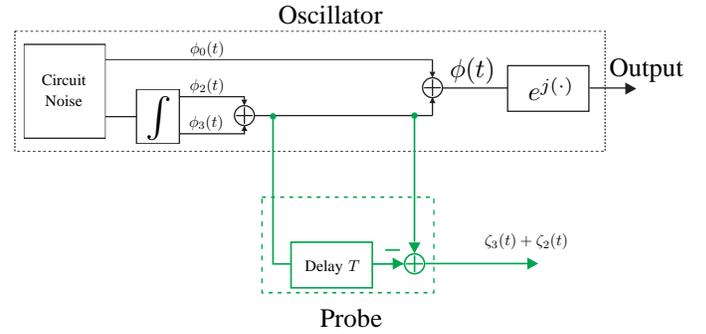}
\caption{Oscillator's internal phase noise generation model.}
\label{fig:3}
\end{figure}

Due to the integration, $\phi_{3}(t)$ and $\phi_{2}(t)$  have an \emph{cumulative} nature \cite{Demir2006,Chorti2006}. PN accumulation over the time delay $T$ can be modeled as the increment phase process 
\begin{subequations}
\label{increment_phase_process}
\begin{align}
\zeta_2(t,T)=\phi_2(t)-\phi_2(t-T)=\int_{t-T}^{t}\Phi_2(\tau)\tr{d}\tau,\\
\zeta_3(t,T)=\phi_3(t)-\phi_3(t-T)=\int_{t-T}^{t}\Phi_3(\tau)\tr{d}\tau,
\end{align}
\end{subequations}
that has been called self-referenced PN \cite{McNeill_thesis}, or the differential PN process \cite{Murat1993} in the literature and it is shown that this process can be accurately modeled as a zero-mean Gaussian process (Fig.~\ref{fig:3}).

\subsection {Communication System Model}
Consider a single carrier communication system. The transmitted signal $x(t)$ is 
\begin{align}
\label{transmitted_cont}
x(t)=\sum_{n=1}^N s[n] p(t-nT),
\end{align}
where $s[n]$ denotes the modulated symbol from constellation $\mathcal{C}$ with average symbol energy of $E_s$, $n$ is the transmitted symbol index, $p(t)$ is a bandlimited square-root Nyquist shaping pulse function with unit-energy, and $T$ is the symbol duration \cite{Proakis2008}.
The continuous-time complex-valued baseband received signal after downconversion, affected by the oscillator PN, can be written as
\begin{align}
\label{Received_cont}
r(t)=x(t)e^{j\phi(t)}+\tilde {w}(t),
\end{align} 
where $\phi(t)$  is the oscillator PN modeled in Sec.~\ref{ssec_PhaseNoiseModel} and $\tilde{w}(t)$ is zero-mean circularly symmetric complex-valued additive white Gaussian noise (AWGN), that models the effect of noise from other components of the system.  The received signal (\ref{Received_cont}) is passed through a matched filter $p^*(-t)$ and the output is
\begin{align}
\label{MF_Output}
y(t)&=\int_{-\infty}^{\infty} \sum_{n=1}^N s[n] p(t-nT-\tau)  p^*(-\tau)e^{j\phi(t-\tau)}\tr{d}\tau\nonumber \\&\hspace{1cm}+\int_{-\infty}^{\infty}\tilde{w}(t-\tau) p^*(-\tau)\tr{d}\tau.
\end{align}
Assuming PN does not change over the symbol duration, but changes from one symbol to another so that no intersymbol
interference arises\footnote{The discrete Wiener PN model, which is well studied in the literature is motivated by this assumption (e.g., \cite{Meyr1997, Tomba98, Armada98, Armada2001, Amblard2003151,Munier2003,Songping2004, Colavolpe2005, Munier2008, Nissila2009, Krishnan2011_1,Krishnan2012_1,Mehrpouyan2012, durisi2013capacity,  Krishnan2013_1,Khanzadi2013_0_COMP}). We also refer the reader to the recent studies of this model where the PN variations over the symbol period has also been taken into consideration, and the loss due to the slowly varying PN approximation has been investigated \cite{martalo2013, ghozlan2013wiener, ghozlan2013multi}.}, sampling the matched filter output (\ref{MF_Output}) at $nT$ time instances results in 
\begin{align}
\label{System_Model1_contTodig}
y(nT)&=s[n]e^{j\phi(nT)}+{w}(nT),
\end{align} 
that with a change in notation we have 
\begin{align}
\label{System_Model1}
y[n]&=s[n]e^{j\phi[n]}+{w}[n],
\end{align} 
where $\phi[n]$ represents the PN of the $n^{th}$ received symbol in digital domain that is bandlimitted after the matched filter, and ${w}[n]$ is the filtered (bandlimitted) and sampled version of $\tilde{w}(t)$ that is a zero-mean circularly symmetric complex-valued AWGN with variance $\sigma^2_w$. Note that in this work our focus is on oscillator phase synchronization and other synchronization issues, such as time synchronization, are assumed perfect.

\begin{figure}[t]
\centering
\psfrag{en}[cc][][0.55]{Encoder}%
\psfrag{dec}[cc][][0.55]{Detector}%
\psfrag{ejp}[cc][][0.8]{$e^{j\phi[n]}$}%
\psfrag{e-jp}[cc][][0.8]{$e^{-j\hat{\phi}[n]}$}%
\psfrag{A1}[cc][][0.6]{AWGN}%
\psfrag{MOD}[cc][][0.51]{Modulator}
\psfrag{DEM}[cc][][0.51]{Demodulator}%
\psfrag{MPSK}[cc][][0.5]{}
\psfrag{OSC}[cc][][0.51]{Oscillator}
\psfrag{w}[cc][][0.8]{$w[n]$}
\psfrag{ES}[cc][][0.49]{ESTIMATOR}
\psfrag{sk}[cc][][0.8]{$s[n]$}
\psfrag{rk}[cc][][0.8]{$y[n]$}
\psfrag{skh}[cc][][0.8]{$\hat{s}[n]$}
\includegraphics [height=1.2in]{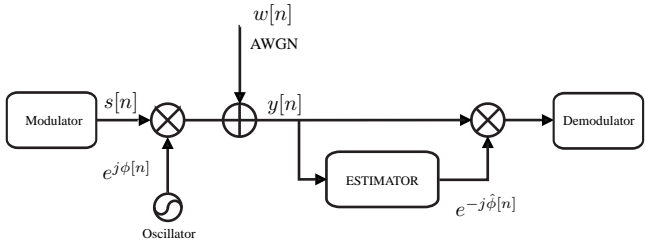}
%%%%\vspace{0.2in}
\caption{Communication system model with a feedforward carrier phase synchronizer \cite{Meyr1997}.}
\label{fig:1}
\end{figure}

\section{System Performance}
\label{Sec_SystemPerformance}
In this section, we find the performance of the introduced communication system from the PN spectrum measurements. Our final result is in terms of error vector magnitude (EVM), which is a commonly used metric for quantifying the accuracy of the received signal \cite{Hassun97,Nakagawa2000}. As shown in Fig.~\ref{fig:1}, PN is estimated at the receiver by passing the received signal through a PN estimator. The estimated PN, denoted as $\hat{\phi}[n]$, is used to de-rotate the received signal before demodulation. The final EVM depends on the accuracy of the PN estimation. In the sequel, we present a bound on the performance of PN estimation, based on the statistics of the PN. 

\subsection{Background: Cram\'{e}r-Rao bounds}
\label{SS_MBCRB}

In order to assess the estimation performance, Cram\'{e}r-Rao bounds (CRBs) can be utilized to give a lower bound on mean square error (MSE) of estimation \cite{book_kay_est}. In case of random parameter estimation, e.g., PN estimation, the Bayesian Cram\'{e}r-Rao bound (BCRB) gives a tight lower bound on the MSE \cite{VanTrees68}. Consider a burst-transmission system, where a sequence of $N$ symbols $\mathbf{s}=[s[1],\dots,s[N]]^T$ is transmitted in each burst.  According to our system model (\ref{System_Model1}), a frame of signals $\mathbf{y}=[y[1],\dots,y[N]]^T$ is received at the receiver with the phase distorted by a vector of oscillator PN denoted as $\boldsymbol{\varphi}=[\phi[1],\dots,\phi[N]]^T$, with the probability density function $f(\boldsymbol{\varphi})$. The BCRB satisfies the following inequality over the MSE of PN estimation:
\begin{align}
\label{BCRB_def}
&\mathbb{E}_{\mathbf{y},\boldsymbol{\varphi}}\left[\left(\hat{\boldsymbol{\varphi}}-\boldsymbol{\varphi}\right)\left(\hat{\boldsymbol{\varphi}}-\boldsymbol{\varphi}\right)^T\right]\geq \mathbf{B}^{-1},\nonumber\\
&\mathbf{B}=\mathbb{E}_{\boldsymbol{\varphi}}\left[\mathbf{F}(\boldsymbol{\varphi})\right]+\mathbb{E}_{\boldsymbol{\varphi}}\left[-\nabla^2_{\boldsymbol{\varphi}} \log f(\boldsymbol{\varphi})\right],
\end{align}
where $\hat{\boldsymbol{\varphi}}$ denotes an estimator of $\boldsymbol{\varphi}$, $\mathbf{B}$ is the Bayesian information matrix (BIM) and ``$\geq$" should be interpreted as meaning that $\mathbb{E}_{\mathbf{y},\boldsymbol{\varphi}}\left[\left(\hat{\boldsymbol{\varphi}}-\boldsymbol{\varphi}\right)\left(\hat{\boldsymbol{\varphi}}-\boldsymbol{\varphi}\right)^T\right]- \mathbf{B}^{-1}$ is positive semi-definite. Here, $\mathbf{F}(\boldsymbol{\varphi})$ is defined as
\begin{align}
\label{FIM_MBCRB}
&\mathbf{F}(\boldsymbol{\varphi})=\mathbb{E}_{\mathbf{s}}\left[\mathbb{E}_{\mathbf{y}|\boldsymbol{\varphi},\mathbf{s}}\left[-\nabla^2_{\boldsymbol{\varphi}} \log f(\mathbf{y}|\boldsymbol{\varphi},\mathbf{s})\right]\right],
\end{align}
and it is called modified Fisher information matrix (FIM) in the literature, and bound calculated from (\ref{BCRB_def}) is equivalently called the modified Bayesian Cram\'{e}r-Rao bound (MBCRB) \cite{Andrea1994}. Based on the definition of the bound in (\ref{BCRB_def}), the diagonal elements of $\mathbf{B}^{-1}$ bound the variance of estimation error of the elements of vector $\boldsymbol{\varphi}$
\begin{align}
\label{error_def}
\sigma^2_{\varepsilon}[n]\triangleq&\mathbb{E}\Big[(\underbrace{{\phi[n]}-\hat{{\phi}}[n]}_{\triangleq\varepsilon[n]})^2\Big]\geq \left[\mathbf{B}^{-1}\right]_{n,n}.
\end{align}

From (\ref{BCRB_def})-(\ref{error_def}), we note that the estimation error variance is entirely determined by the prior probability density function (pdf) of the PN $f(\boldsymbol{\varphi})$ and the conditional pdf of the received signal $\mathbf{y}$ given the PN and transmitted signal $f(\mathbf{y}|\boldsymbol{\varphi},\mathbf{s})$ (usually denoted as the likelihood of $\boldsymbol{\varphi}$). In the following, we derive those pdfs based on our models in Sec.~\ref{sec_system} and use them in our calculations.
\subsection{Calculation of the bound}
\subsubsection{Calculation of $\mathbb{E}_{\boldsymbol{\varphi}}\left[-\nabla^2_{\boldsymbol{\varphi}} \log f(\boldsymbol{\varphi})\right]$} 
Based on our PN model (\ref{PN_parts}) and the phase increment process defined in (\ref{increment_phase_process}), the sampled PN after the matched filter can be written as
\begin{align} 
\label{phase_increament_dis2}
\phi[n]&=\phi_3[n]+\phi_2[n]+\phi_0[n],\nonumber\\
%&=\sum_{i=1}^{n}\zeta_3[i]+\sum_{i=1}^{n}\zeta_2[i]+\phi_0[n],\\
&=\phi_3[1]+\sum_{i=2}^{n}\zeta_3[i]+\phi_2[1]+\sum_{i=2}^{n}\zeta_2[i]+\phi_0[n],
\end{align}
where $\zeta_3[n]\triangleq\zeta_3(nT,T)$ and $\zeta_2[n]\triangleq\zeta_2(nT,T)$ are the discrete-time phase increment processes, and  $\phi_3[1]$ and $\phi_2[1]$ are the cumulative PN of the first symbol in the block, which are modeled as zero-mean Gaussian random variables with a high variance\footnote{We consider a flat non-informative prior \cite{book_kay_est,Bay2008} for the initial PN values. To simplify the derivations, it is modeled by a Gaussian distribution with a high variance that is wrapped to a flat prior over $[0,2\pi]$.}, denoted as $\sigma^2_{\phi_3[1]}$ and $\sigma^2_{\phi_2[1]}$, respectively. According to (\ref{phase_increament_dis2}) and due to the fact that $\zeta_3[n]$ and $\zeta_2[n]$ are samples from zero-mean Gaussian random processes, $\boldsymbol{\varphi}$ has a zero-mean multivariate Gaussian prior $f(\boldsymbol{\varphi})=\mathcal{N}(\boldsymbol{\varphi};\mathbf{0},\mathbf{C})$, where $\mathbf{C}$ denotes the covariance matrix whose elements are computed in Appendix~\ref{Appendix_B} as
\begin{align}
\label{CovMatrix}
%\phi_\tr{c}(1)+\sum_{i=1}^{n-1}\zeta(i)+\phi_0[n]\\
[\mathbf{C}]_{l,k}=&\underbrace{\sigma^2_{\phi_3[1]}+\sum_{m=2}^{l}\sum_{m'=2}^{k}R_{\zeta_3}[m-m']}_{\tr{from}~\phi_3[n]}\nonumber\\
&+\underbrace{\sigma^2_{\phi_2[1]}+\sum_{m=2}^{l}\sum_{m'=2}^{k}R_{\zeta_2}[m-m']}_{\tr{from}~\phi_2[n]}\nonumber\\
&+\underbrace{\delta[l-k]\sigma_{\phi_0}^2}_{\tr{from}~\phi_0[n]},\nonumber\\
&\hspace{-1.1cm}l,k=\{1\dots N\},
\end{align}
where $R_{\zeta_3}[m]$ and $R_{\zeta_2}[m]$ are the autocorrelation functions of $\zeta_3[n]$ and $\zeta_2[n]$, and $\sigma_{\phi_0}^2$ is the variance of $\phi_0[n]$. The required statistics, i.e., $R_{\zeta_3}[m]$, $R_{\zeta_2}[m]$ and $\sigma_{\phi_0}^2$ can be computed from the oscillator PN measurements. To keep the flow of this section, we derive these statistics in Sec.~\ref{sec_PhaseNoiseStatistics}, where the final results are presented in (\ref{WhitePN_Var}), ({\ref{PLL_f1_ACF_sampled}}), (\ref{f1_ACF_sampled}) and (\ref{f3_ACF_Summary}). Finally, based on the definition of $f(\boldsymbol{\varphi})$, it is straightforward to show that $\nabla^2_{\boldsymbol{\varphi}} \log f(\boldsymbol{\varphi})=-\mathbf{C}^{-1}$, and consequently due to the independence of $\mathbf{C}$ from $\boldsymbol{\varphi}$
\begin{align}
\label{Log_Prior_D2}
\mathbb{E}_{\boldsymbol{\varphi}}\left[-\nabla^2_{\boldsymbol{\varphi}} \log f(\boldsymbol{\varphi})\right]=\mathbf{C}^{-1}.
\end{align}

\subsubsection{Calculation of $\mathbb{E}_{\boldsymbol{\varphi}}\left[\mathbf{F}(\boldsymbol{\varphi})\right]$}

According to the system model in (\ref{System_Model1}), the likelihood function is written as 
\begin{align}
\label{DA_Likelihood}
f(\mathbf{y}|\boldsymbol{\varphi},\mathbf{s})&=\prod_{n=1}^Nf\left(y[n]|\phi[n],s[n]\right)\nonumber\\
&=\left(\frac{1}{\sigma_w^2\pi}\right)^N\prod_{n=1}^Ne^{-\frac{|y[n]|^2+|s[n]|^2}{\sigma_w^2}}\nonumber\\
&\hspace{2.1cm}\times e^{\frac{2}{\sigma^2_w}\Re\{y[n] s^*[n]e^{-j\phi[n]}\}},
\end{align}
where the first equality is due to independence of the AWGN samples. We can easily show that $\nabla^2_{\boldsymbol{\varphi}} \log f(\mathbf{y}|\boldsymbol{\varphi},\mathbf{s})$ is a diagonal matrix where its diagonal elements are
\begin{align}
\label{DA_Likelihood_D2}
\left[\nabla^2_{\boldsymbol{\varphi}} \log f(\mathbf{y}|\boldsymbol{\varphi},\mathbf{s})\right]_{n,n}&=\frac{\partial^2 \log f(y[n]|\phi[n],s[n])}{\partial\phi^2[n]}\nonumber\\
&=-\frac{2}{\sigma^2_w}\Re\{y[n] s^*[n]e^{-j\phi[n]}\}.
\end{align}
Following (\ref{FIM_MBCRB}) and (\ref{DA_Likelihood_D2}), diagonal elements of FIM are computed as
\begin{align}
\label{DA_Likelihood_D2_Exp}
\left[\mathbf{F}(\boldsymbol{\varphi})\right]_{n,n}=\frac{2E_s}{\sigma^2_w},
%=\mathbb{E}_{\mathbf{s}}\left[\mathbb{E}_{\mathbf{y}|\boldsymbol{\varphi},\mathbf{s}}\left[\frac{2}{\sigma^2_w}\Re\{y[n] s^*[n]e^{-j\phi[n]}\}\right]\right]\\
\end{align}
where $E_s$ is the average energy of the signal constellation. This implies that 
\begin{align}
\label{FIM_Final}
\mathbf{F}(\boldsymbol{\varphi})=\frac{2E_s}{\sigma^2_w} \mathbf{I},
\end{align}
where $\mathbf{I}$ is the identity matrix. Finally, from (\ref{BCRB_def}), (\ref{Log_Prior_D2}), and (\ref{FIM_Final}) 
\begin{align}
\label{MBCRB_final}
\mathbf{B}=\frac{2E_s}{\sigma^2_w} \mathbf{I}+\mathbf{C}^{-1}.
\end{align}
The minimum MSE of PN estimation (\ref{error_def}) depends on SSB PN spectrum measurements through $\mathbf{B}$ and $\mathbf{C}^{-1}$. We will use this result in the following subsection to calculate a more practical performance measure that is called EVM. 

\subsection{Calculation of Error Vector Magnitude}
The modulation accuracy can be quantified by the EVM, defined as the root-mean square error between the transmitted and received symbols \cite{Hassun97,Nakagawa2000}
\begin{align}
\label{EVM_Def}
\tr{EVM}[n]=\sqrt{\frac{\frac{1}{M}\sum^M_{k=1}|s_k[n]-s'_k[n]|^2}{E_s}},
\end{align}
where $s_k[n]$, $k\in\{1,\dots,M\}$, is the transmitted symbol from the constellation $\mathcal{C}$ with order $M$, at the $n^{th}$ time instance, and $s'_k[n]$ is the distorted signal at the receiver. Even with optimal PN estimators, we have residual phase errors. Hence, cancellation of PN by de-rotation of the received signal with the estimated PN results in a distorted signal
\begin{align}
s'_k[n]&= s_k[n]e^{j(\phi[n]-\hat{\phi}[n])}\nonumber\\
&=s_k[n]e^{j\varepsilon[n]},
\end{align}
where $\varepsilon[n]$ is the residual phase error. Before going further, assume we have used an PN estimator \cite{book_kay_est} that reaches the computed MBCRB, and estimation error $\varepsilon[n]$ is a zero-mean Gaussian random variable. Our numerical evaluations in the result section support the existence of such estimators (Fig.~\ref{fig:6_0}). This implies that $f(\varepsilon[n])=\mathcal{N}(\varepsilon[n];0,\sigma^2_{\varepsilon}[n])$, where $\sigma^2_{\varepsilon}[n]$ is defined in (\ref{error_def}) and can be computed from the derived MBCRB. The variance obtained from the MBCRB results from averaging over all possible transmitted symbols. Note that to calculate the EVM accurately, we need to use the conditional PDF of the residual PN variance $f(\varepsilon[n]|\mathbf{s})$. However, in order to keep our analysis less complex we approximate the conditional PDF with the unconditional one: $f(\varepsilon[n]|\mathbf{s})\approx f(\varepsilon[n])$. Our numerical simulations show the validity of this approximation in several scenarios of interest (Fig.~\ref{fig:6}). For the sake of notational simplicity, we drop the time index $n$ in the following calculations. Averaging over all possible values of $\varepsilon[n]$, (\ref{EVM_Def}) is rewritten as 
\begin{align}
\label{EVM_Def_cont}
\tr{EVM}[n]=\sqrt{\frac{\frac{1}{M}\sum^M_{k=1}\mathbb{E}_{\varepsilon}\Big[|s_k-s_k e^{j\varepsilon}|^2\Big]}{E_s}}.
\end{align}
The magnitude square of the error vector for a given $\varepsilon[n]$, and $s_k[n]$ is determined as 
\begin{align}
|s_k-s_k e^{j\varepsilon}|^2&=2|s_k|^2(1-\cos(\varepsilon))\nonumber
\\&=4|s_k|^2\sin^2(\frac{\varepsilon}{2}),
\end{align}
and consequently 
\begin{align}
\label{Vector_Mag_Calc}
\hspace{-0.25cm}\frac{1}{M}\sum^M_{k=1}\mathbb{E}_{\varepsilon}\Big[|s_k-s_ke^{j\varepsilon}|^2\Big]&=4\frac{1}{M}\sum^M_{k=1}|s_k|^2\mathbb{E}_{\varepsilon}\Big[\sin^2(\frac{\varepsilon}{2})\Big].
\end{align}
The expectation in (\ref{Vector_Mag_Calc}) can be computed as
\begin{align}
\label{Vector_Mag_Calc_integral}
\mathbb{E}_{\varepsilon}\Big[\sin^2(\frac{\varepsilon}{2})\Big]&=
\int_{-\infty}^{\infty} \sin^2(\frac{\varepsilon}{2})f(\varepsilon)\tr{d}\varepsilon \nonumber \\
&=(1-e^{-{\sigma_{\varepsilon}^2}/{2}})/2,
\end{align}
where $f(\varepsilon)$ is the Gaussian pdf of $\varepsilon[n]$ as defined before.
%Finally, EVM can be computed from (\ref{EVM_Def_cont}), (\ref{Vector_Mag_Calc}) and (\ref{Vector_Mag_Calc_integral}) as
%\begin{align}
%\label{EVM_final}
%\tr{EVM}[n]=\sqrt{2(1-e^{-{\sigma_{\varepsilon}^2[n]}/{2}})}.
%\end{align}
%As can be seen in (\ref{EVM_final}), EVM is only a function of $\sigma^2_{\varepsilon}[n]$, i.e., the variance of the residual phase errors. 

Finally, EVM can be computed from (\ref{EVM_Def_cont}), (\ref{Vector_Mag_Calc}), and (\ref{Vector_Mag_Calc_integral}) as
\begin{align}
\label{EVM_final}
\tr{EVM}[n]&=\sqrt{2(1-e^{-{\sigma_{\varepsilon}^2[n]}/{2}})}\nonumber\\
&=\sqrt{2(1-\exp{(\left[-0.5\mathbf{B}^{-1}\right]_{n,n}})}\nonumber\\
&=\sqrt{2-2\exp{\left(-0.5{\left[\left(\frac{2E_s}{\sigma^2_w}\mathbf{I}+\mathbf{C}^{-1}\right)^{-1}\right]}_{n,n}\right)}},
\end{align}
where $\mathbf{C}$ is calculated in (\ref{CovMatrix}) and it is a function of PN model parameters $K_3$, $K_2$, and $K_0$ through $R_{\zeta_3}[m]$, $R_{\zeta_2}[m]$ and $\sigma^2_{\phi_{0}}$, computed in Sec.~\ref{sec_PhaseNoiseStatistics}.

\section{Phase Noise Statistics}
\label{sec_PhaseNoiseStatistics}
As we can see in Sec.~\ref{Sec_SystemPerformance}, the final system performance computed in terms of EVM (\ref{EVM_final}) depends on the minimum MSE of the PN estimation defined in (\ref{error_def}). According to (\ref{CovMatrix}), in order to find the minimum PN variance, we have to compute the required PN statistics; i.e., $R_{\zeta_3}[m]$, $R_{\zeta_2}[m]$ and $\sigma_{\phi_0}^2$. To find these statistics we need to start from our continuous-time PN model described in Sec.~\ref{ssec_PhaseNoiseModel}. Based on (\ref{increment_phase_process}), $\phi_3(t)$ and $\phi_2(t)$ result from integration of noise sources inside the oscillator. On the other hand, $\phi_0(t)$ has external sources. Therefore, we separately study the statistics of these two parts of the PN.
\subsection{Calculation of $\sigma_{\phi_0}^2$}
The PSD of $\phi_{0}(t)$ is defined as 
\begin{align}
\label{f0_PSD}
S_{\phi_0}(f)=K_0,
\end{align}
where $K_0$ is the level of the noise floor that can be found from the measurements, and according to (\ref{normalized_spectrum}), it is normalized with the oscillator power \cite{Leeson1966}. The system bandwidth is equal to the symbol rate\footnote{This bandwidth corresponds to using a raised-cosine pulse shaping filter $p(t)$ defined in (\ref{transmitted_cont}) with zero excess bandwidth. For the general case, the bandwidth becomes ${(1+\alpha)}/{T}$ where $\alpha$ denotes the excess bandwidth \cite{Proakis2008}.} $1/T$, and at the receiver, a low-pass filter with the same bandwidth is applied to the received signal $x(t)e^{j\phi_0(t)}$. According to \cite{Corvaja2002}, if $K_0/T$ is small (which is generally the case in practice), low-pass filtering of the received signal results in filtering of $\phi_0(t)$ with the same bandwidth. Therefore we are interested in the part of the PN process inside the system bandwidth. The variance of the bandlimited $\phi_{0}(t)$ is calculated as 
\begin{align}
\label{WhitePN_Var}
\sigma^2_{\phi_{0}}=\int^{+1/2T}_{-1/2T}S_{\phi_0}(f)\tr{d}f=\frac{K_0}{T}.
\end{align}
As $\phi_{0}(t)$ is bandlimited, we can sample it without any aliasing. 

\subsection{Calculation of $R_{\zeta_3}[m]$ and $R_{\zeta_2}[m]$}
It is possible to show that $\zeta_3(t,T)$ and $\zeta_2(t,T)$ defined in (\ref{increment_phase_process}) are stationary processes and their variance over the time delay $T$, is proportional to $T$ and  $T^2$, respectively \cite{McNeill1997,Liu2004,Demir2006}. However, as shown in this work, their variance is not enough to judge the effect of using a noisy oscillator on the performance of a communication system, and hence their autocorrelation functions must be also taken into consideration. 
%However, in order to evaluate the system performance, we need to study the statistical properties of JAP resulted from the cumulative PN in more detail.
%Accumulation of the cumulative part of PN, denoted as $\phi_\tr{c}(t)\triangleq\phi_{2}(t)+\phi_{3}(t)$, over the time can be modeled as a RP that in this work is called jitter accumulation process (JAP). It is possible to show that the variance of the JAP, introduced over the time delay $T$ by $\phi_{2}(t)$ and $\phi_{3}(t)$, is proportional to $T$ and  $T^2$, respectively \cite{Liu2004,Demir2006}. However, in order to evaluate the system performance, we need to study the statistical properties of JAP resulted from the cumulative PN in more detail. The continuous-time JAP  $\zeta(t)$ is defined as accumulation process of the phase noise over the time delay $T$ 
%\begin{align} 
%\label{phase_increament_cont}
%\zeta(t,T)=\phi_\tr{c}(t)-\phi_\tr{c}(t-T).
%\end{align}
%JAP is a stationary RP with zero-mean Gaussian distribution and as discussed in \cite{Liu2004}, the procedure of finding the samples of this process can be expressed by a linear time invariant (LTI) sampling system with impulse response of
Samples of $\zeta_3(t,T)$ and $\zeta_2(t,T)$ can be found by applying a delay-difference operator on $\phi_3(t)$ and $\phi_2(t)$, respectively \cite{Liu2004,Demir2006}, which is a linear time invariant sampling system with impulse response of
\begin{align} 
\label{impulse_response_h}
h(t)=\delta(t)-\delta(t-T).
\end{align}
Starting from $\phi_2(t)$, the PSD of $\zeta_2(t,T)$ can be computed as
\begin{align} 
\label{JAP_PSD}
S_{\zeta_{2}}(f)=S_{\phi_{2}}(f)|H(j2\pi f)|^2,
\end{align}
where $H(j2\pi f)=1-e^{-j2\pi f T}$ is the frequency response of the delay-difference operator introduced in (\ref{impulse_response_h}). 
The autocorrelation function of $\zeta_2(t,T)$ can be computed by taking the inverse Fourier transform of its PSD 
\begin{align} 
\label{General_ACF}
R_{\zeta_{2}}(\tau)=\int^{+\infty}_{-\infty}S_{\zeta_{2}}(f) e^{j2\pi f \tau}\tr{d}f,
\end{align}
where $\tau$ is the time lag parameter. Using (\ref{JAP_PSD}) and (\ref{General_ACF}) the continuous-time auto correlation function can be found as
\begin{align}
\label{Main_Cont_ACF}
R_{\zeta_{2}}(\tau)=8\int^{+\infty}_{0}S_{\phi_{2}}(f)\sin(\pi f T)^2 \cos(2\pi f \tau)\tr{d}f.
\end{align}
As can be seen in (\ref{Main_Cont_ACF}), in order to find the closed-form autocorrelation functions, we do not confine our calculations inside the system bandwidth $1/T$. However, we see from the measurements that parts of $\phi_{3}(t)$ and $\phi_{2}(t)$ outside bandwidth are almost negligible and do not have any significant effect on the calculated autocorrelation functions.

The PSD of $\phi_{2}(t)$ has the form of
\begin{align}
\label{f2_PSD}
S_{\phi_{2}}(f)=\frac{K_2}{f^2+\gamma^2},
\end{align}
where $K_2$ and can be found from the measurements and $\gamma$ is a low cut-off frequency that is considered to be very small for a free running oscillator, while it is set to the PLL's loop bandwidth in case of using a locked oscillator  (Fig.~\ref{fig:2}-b). 
According to (\ref{Main_Cont_ACF}) and (\ref{f2_PSD}), autocorrelation function of $\zeta_2(t,T)$ can be determined as
\begin{align}
\label{f1_ACF}
\hspace{-0.5cm}R_{\zeta_{2}}(\tau)&=8\int^{+\infty}_{0}\frac{K_2}{f^2+\gamma^2}\sin(\pi fT)^2 \cos( 2\pi f \tau)\tr{d}f\nonumber\\
%&=2K_2\pi^2 \left(-2|\tau|+|\tau-T|+|\tau+T|\right).
&=\frac{K_2\pi}{\gamma} \left(2 e^{-2\gamma\pi|\tau|}-e^{-2\gamma\pi|\tau-T|}-e^{-2\gamma\pi|\tau+T|}\right).
\hspace{-0.4cm}
\end{align}
Sampling (\ref{f1_ACF}) results in
\begin{align}
\label{PLL_f1_ACF_sampled}
R_{\zeta_{2}}[m]=\frac{K_2\pi}{\gamma} \left(2 e^{-2\gamma\pi T |m|}-e^{-2\gamma\pi T|m-1|}-e^{-2\gamma\pi T|m+1|}\right),
\end{align}
where $R_{\zeta_{2}}[m]\triangleq R_{\zeta_{2}}(mT)$. For a free running oscillator, the autocorrelation function can be found by taking the limit of ({\ref{PLL_f1_ACF_sampled}}) as $\gamma$ approaches $0$, that results in
\begin{align}
\label{f1_ACF_sampled}
R_{\zeta_{2}}[m]
=\begin{cases}
   4K_2\pi^2T & \text{if } m = 0 \\
   0 & \text{otherwise }\\
  \end{cases}.
\end{align}
Results in ({\ref{PLL_f1_ACF_sampled}}) and ({\ref{f1_ACF_sampled}}) show that for a locked oscillator $\zeta_2[n]$ is a colored process (its samples are correlated with each other), while it is white for a free running oscillator.
\begin{figure}[t]
\centering
\psfrag{w0}[cc][][0.9]{$w_0[n]$}
\psfrag{w3}[cc][][0.9]{$w_3[n]$}%
\psfrag{w2}[cc][][0.9]{$w_2[n]$}%
\psfrag{p0}[cc][][0.9]{$\phi_0[n]$}
\psfrag{p3}[cc][][0.9]{$\phi_3[n]$}%
\psfrag{p2}[cc][][0.9]{$\phi_2[n]$}%
\psfrag{tot}[cc][][1]{$\phi[n]$}
\psfrag{H0}[cc][][0.75]{$H_0(z)=1$}%
\psfrag{H2}[cc][][0.75]{$H_2(z)=\frac{1}{(1-z^{-1})}$}%
\psfrag{H3}[cc][][0.75]{$H_3(z)=\frac{1}{(1-z^{-1})^{3/2}}$}%
\includegraphics [width=3.2in]{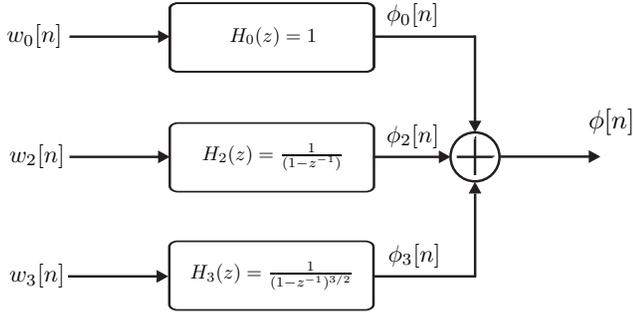}
\caption{Phase noise sample generator.}
\label{fig:4}
\end{figure}
To find $R_{\zeta_{3}}[m]$ for a free running oscillator, one can consider the PSD of $\phi_{3}(t)$ to be $S_{\phi_{3}}(f)\propto 1/f^3$. However, by doing so, $S_{\zeta_{3}}(f)$ defined  in (\ref{JAP_PSD}) diverges to infinity at zero offset frequency and hence makes it impossible to find the autocorrelation function in this case. To resolve the divergence problem, we follow a similar approach to \cite{Liu2004,Demir2006} and introduce a low cutoff frequency $\gamma$ below which $S_{\phi_{3}}(f)$ flattens. Our numerical studies show that as long as $\gamma$ is chosen reasonably small, its value does not have any significant effect on the final result. Similar to our analysis for $\phi_{2}$, the autocorrelation of PN increments at the output of a first order PLL can be found by setting $\gamma$ equal to the PLL's loop bandwidth. Hence, we define the PSD of $\phi_{3}(t)$ as
\begin{align}
\label{f3_PSD}
S_{\phi_{3}}(f)=\frac{K_3}{|f|^3+\gamma^3},
\end{align}
where $K_3$ can be found from the measurements (Fig.~\ref{fig:1}). Following the same procedure of calculating $R_{\zeta_{2}}(\tau)$ in (\ref{JAP_PSD}-\ref{Main_Cont_ACF}) and using (\ref{f3_PSD}), the autocorrelation function of $\zeta_{3}(t)$ can be computed by solving the following integral
\begin{align}
\label{f3_PSD_Cont}
R_{\zeta_{3}}(\tau)=8\int^{+\infty}_{0}\frac{K_3}{f^3+\gamma^3}\sin(\pi fT)^2 \cos( 2\pi f \tau)\tr{d}f.
\end{align}
This integral is solved in the Appendix~\ref{Appendix_A}. Finally, the closed-form sampled autocorrelation function of $\zeta_3[n]$ is approximated as
\begin{subequations}
\label{f3_ACF_Summary}
\begin{align}
&R_{\zeta_3}[0]\approx-8K_3\pi^2 T^2 \left(\Lambda+\log(2\pi\gamma T)\right)\\
&R_{\zeta_3}[\pm 1]\approx-8K_3\pi^2  T^2 (\Lambda+\log(8\pi\gamma T)),
\end{align}
otherwise
\begin{align}
R_{\zeta_3}[m]
\approx&-8K_3\pi^2T^2 \Big[-m^2(\Lambda+\log(2\pi\gamma T|m|))\nonumber\\
&+\frac{(m+1)^2}{2}(\Lambda+\log(2\pi\gamma T|m+1|))\nonumber\\
&+\frac{(m-1)^2}{2}(\Lambda+\log(2\pi\gamma T|m-1|))\Big],
\end{align}
\end{subequations}
where $\Lambda\triangleq\Gamma-{3}/{2}$, and $\Gamma\approx 0.5772$ is the Euler-Mascheroni's constant \cite{havil2003gamma}. The calculated variance $R_{\zeta_{3}}[0]$ is almost proportional to $T^2$ which is similar to the results of \cite{Liu2004,Demir2006}. As it can be seen from (\ref{f3_ACF_Summary}), samples of $\zeta_{3}[n]$ are correlated in this case which is in contrast to $\zeta_{2}[n]$. Consequently, in presence of $\phi_3(t)$, variance of $\zeta_3[n]$ is not adequate to judge the behavior of the oscillator in a system; it is necessary to incorporate the correlation properties of $\zeta_3[n]$ samples.
\section{Numerical and Measurement Results}
\label{Sec_NumericalSimulation}
In this section, first the analytical results obtained in the previous sections are evaluated by performing Monte-Carlo simulations. Then, the proposed EVM bound is used to quantify the system performance for a given SSB PN measurement.
\subsection{Phase Noise Simulation}
\label{Phase_Noise_Simulation}
To evaluate our proposed EVM bound, we first study the generation of time-domain samples of PN that match a given PN SSB measurement in the frequency domain. As shown in (\ref{PN_parts}), we model PN as a summation of three independent noise processes $\phi_{3}(t)$, $\phi_{2}(t)$, and $\phi_{0}(t)$. The same model is followed to generate time-domain samples of the total PN process (Fig.~\ref{fig:4}). Generating the samples of power-law noise with PSD of $1/f^\alpha$ has been vastly studied in the literature \cite{Kasdin1995,Yousefi2010Model,Khanzadi2011}. One suggested approach in \cite{Kasdin1995} is to pass independent identically distributed (iid) samples of a discrete-time Gaussian noise process through a linear filter with the impulse response of 
\begin{align}
\label{PN_filter}
H(z)=\frac{1}{(1-z^{-1})^{\alpha/2}}.
\end{align} 
The PSD of the generated noise can be computed as 
\begin{align}
\label{disc_PSD}
S^d(f)=\sigma^2_{w_\alpha}H(z)H(z^{-1})T,
\end{align}
where $T$ is the sampling time equal to the symbol duration, and $\sigma^2_{w_\alpha}$ is the variance of input iid Gaussian noise \cite{Kasdin1995}. Fig.~\ref{fig:4} illustrates the block diagram used for generating the total PN process. Tab.~\ref{tab:PN_Generation} shows variance of the input iid Gaussian noise in each branch calculated based on (\ref{disc_PSD}).
\begin{table}[h]
\renewcommand{\arraystretch}{1.5}
\caption{PN generation: Input iid Noise Variance}
\label{tab:PN_Generation}
\centering
\begin{tabular}{| c | c | l |}
  \hline                        
  PN Process & PSD & Input Noise Variance \\
\hline\hline
$\phi_0[n]$ & $K_0$ & $\quad \sigma^2_{w_0}={K_0}/{T}$\\
\hline
$\phi_2[n]$ & $K_2/f^2$ & $\quad \sigma^2_{w_2}=4 K_2 T \pi^2$\\
\hline  
$\phi_3[n]$ & $K_3/f^3$ & $\quad \sigma^2_{w_3}=8 K_3 T^{2} \pi^3$\\
\hline  
\end{tabular}
\end{table}
\begin{figure}[t]
\centering
\psfrag{Lk0}[cc][][0.8]{{\color[rgb]{0,0,1}$-110~\tr{dBc/Hz}$}}%
\psfrag{Lk2}[cc][][1]{$\frac{10}{f^2}$}%
\psfrag{Lk3}[cc][][1]{{\color[rgb]{1,0,0}$\frac{10^4}{f^3}$}}%
\psfrag{w0}[cc][][1]{$w_0[n]$}%
\includegraphics [width=3.4in,height=2.72in]{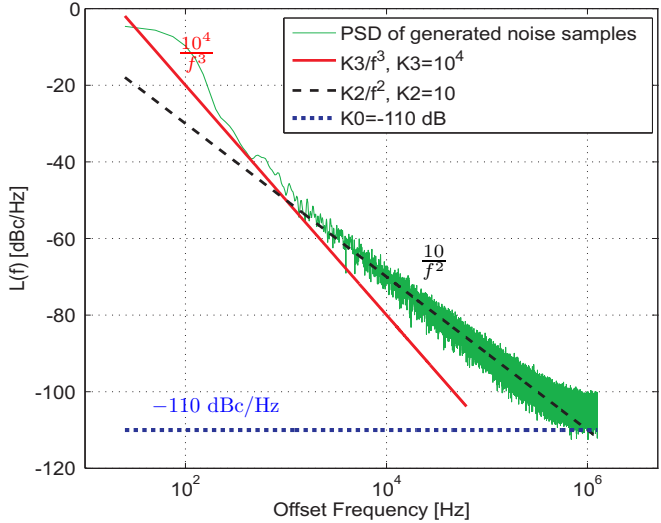}
\caption{PSD of the generated PN samples vs. the theoretical PSD. The generated phase noise PSD is matched to the desired PSD with the given values of $K_0$, $K_1$ and $K_3$.}
\label{fig:5}
\end{figure}
Fig.~\ref{fig:5} shows the total one-sided PSD of the generated PN samples for a particular example. The frequency figures of merits are set to be $K_0=-110~\tr{dB}$, $K_3=10^4$, and $K_2=10$. According to Tab.~\ref{tab:PN_Generation}, the variance of input white Gaussian noises to the PN generation system in Fig.~\ref{fig:4} for a system with symbol rate $10^6$~symbol/sec are calculated to be $\sigma^2_{w_0}=5\times 10^{-6}$, $\sigma^2_{w_2}=1.97 \times 10^{-4}$, and $\sigma^2_{w_3}=1.26\times 10^{-6}$. This figure shows that generated time-domain samples match to the PSD of PN. 
%It can also be seen that the PSD of the simulated PN flattens below certain offset frequencies and it has a Gaussian shape. We refer the reader to \cite{Klimovitch2000_1,Klimovitch2000_2,Yousefi2010Model} for more details on mathematical explanation of this phenomenon. 

\subsection {Monte-Carlo Simulation}
\begin{figure}[t]
\centering
\includegraphics [height=2.9in]{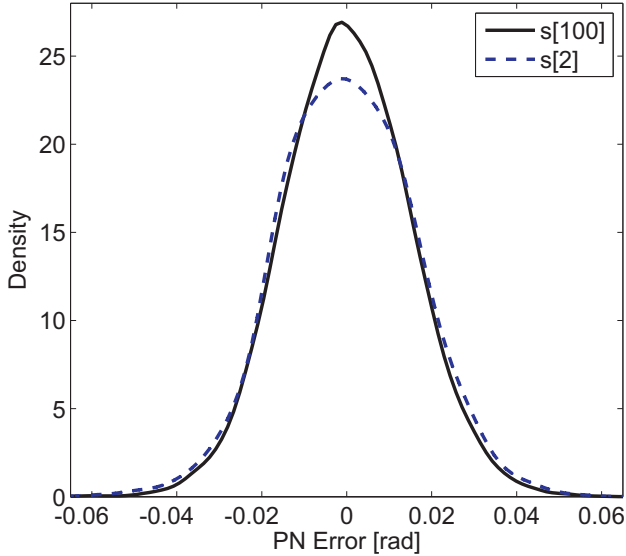}
\caption{The phase error distribution of the second symbol $(n=2)$ and mid symbol of the block $(n=100)$ estimated from $10000$ simulation trials. It can be seen that the phase error distribution is almost zero-mean Gaussian for both symbols. PN of the symbol in the middle of the block can be estimated better and has a lower residual variance.}
\label{fig:6_0}
\end{figure}
\begin{figure}[t]
\centering
\includegraphics [height=2.7in]{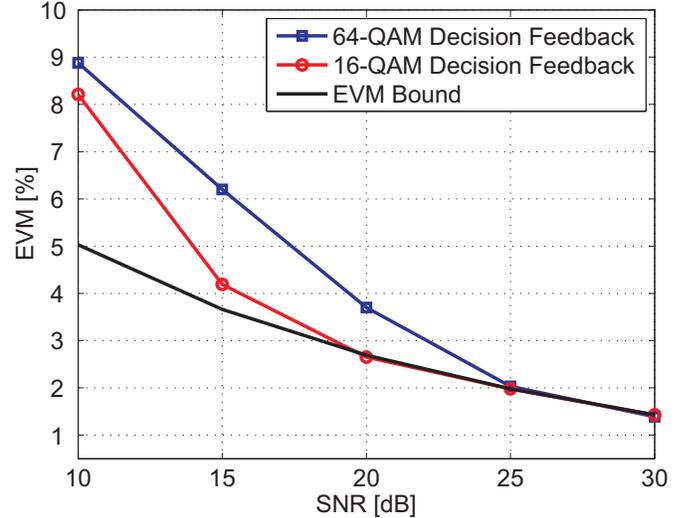}
\caption{Proposed theoretical EVM bound vs. the EVM from the Monte-Carlo simulation. The PSD in Fig.~\ref{fig:5} is considered as the PN PSD. 16 and 64-QAM modulations are used, pilot density is $10\%$, and the symbol rate is set to $10^6~\tr{[Symbol/sec]}$. Note that in pure AWGN case, the symbol error probability of 16-QAM at SNR$=20$~dB is $10^{-5}$ and for 64-QAM it is $10^{-4}$ at SNR=$25$~dB.}
\label{fig:6}
\end{figure}
Consider a communication system like that of Fig.~\ref{fig:1}. Two modulation schemes i.e., 16-QAM and 64-QAM are used and length of the communication block is set to 200 symbols. A local oscillator with the PN PSD of Fig.~\ref{fig:5} is used. For the Monte-Carlo simulation, we first generate the PN samples following the routine proposed in Sec.~\ref{Phase_Noise_Simulation}. Then, we design the maximum a posteriori (MAP) estimator of the PN vector ${\boldsymbol{\varphi}}$ at the receiver. The MAP estimator is a Bayesian estimator that can be used for estimation of random parameters \cite{VanTrees68,book_kay_est}. This estimator finds $\hat{\boldsymbol{\varphi}}$ that maximizes the posteriori distribution of ${\boldsymbol{\varphi}}$: 
\begin{align}
\label{MAP_def}
\boldsymbol {\hat{\varphi}}_\tr{MAP}&=\underset{\boldsymbol {\varphi}}{\arg \max}~ 
f(\boldsymbol {\varphi}|\mathbf{y},\mathbf {s})\nonumber\\&=\underset{\boldsymbol {\varphi}}{\arg \max}~f(\mathbf{y}|\boldsymbol {\varphi},\mathbf {s})f(\boldsymbol {\varphi}).
%f(\boldsymbol {\theta}|\mathbf{y},\mathbf {s})=f(\mathbf{y}|\boldsymbol {\theta},\mathbf {s})f(\boldsymbol {\theta})
\end{align}
The needed likelihood and prior functions for designing this estimator are calculated in Sec.~\ref{Sec_SystemPerformance}. However, the detailed implementation of this estimator is not in the focus of this paper and we focus only on the final results. We refer the interested reader to \cite{Khanzadi2013_2_ColoredPNEst,VanTrees68,book_kay_est,Amblard2003151} for more information on implementation of the MAP and other Bayesian estimators such as Kalman or particle filters, that can be used for estimation of random parameters. The estimated phase values from the MAP estimator are used to eliminate the effect of PN by de-rotation of the received signals. Finally, the EVM is computed by comparing the transmitted symbols with the signal after PN compensation. 
Fig.~\ref{fig:6_0} shows the density of the residual phase errors for two of the symbols in the frame ($n=2$ and $n=100$). It can be seen that phase errors are almost zero mean and have Gaussian distribution. PN in the middle of the block can be estimated better has a lower residual variance. Fig.~\ref{fig:6} compares the proposed theoretical EVM bound (average EVM over the block) against the resulted EVM calculated from the Monte-Carlo simulation of a practical system. In this simulation, 16-QAM and 64-QAM modulations are used, where $10\%$ of the symbols are known (pilot symbols) at the receiver. For the unknown symbols, decision-feedback from a symbol detector is used at the estimator. It can be seen that the calculated EVM from the empirical simulation matches the proposed theoretical bound at moderate and high SNRs. It can also be seen that at low SNR, the bound is more accurate for 16-QAM modulation format. This is mainly due to the fact that $16$-QAM has a lower symbol error probability than $64$-QAM for a given SNR, thus the decision-feedback is more accurate in this case. 
\subsection{Analysis of the Results}
\label{SS_Discussions}
\begin{figure}[t]
\centering
\psfrag{l-cf}[cc][][0.8]{$\mathcal{L}(f)$}%
\psfrag{l-f}[cc][][0.8]{$f$}%
\includegraphics [width=3.55in]{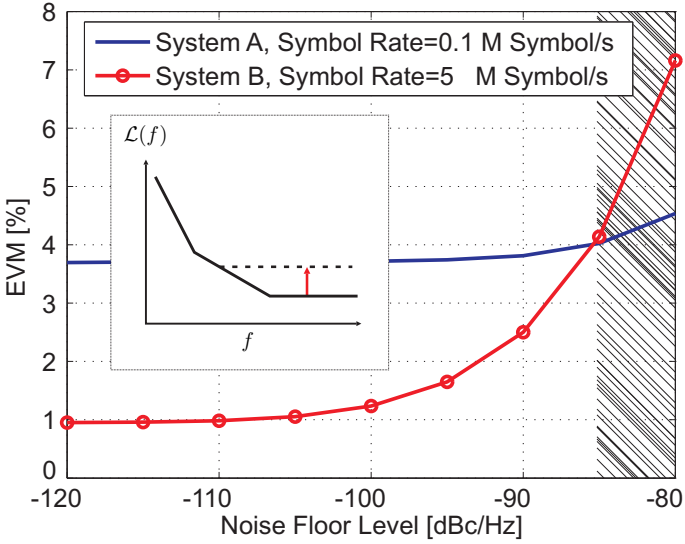}
\caption{The proposed theoretical EVM bound against different noise floor levels. $K_2=1$ and $K_3=10^4$ are kept constant. The low cut-off frequency $\gamma$ is considered to be $1~\tr{Hz}$ and SNR$=30$~dB and block-length is set to $10$. In the hatched regime, the white PN (noise floor) dominates over the cumulative part of the PN.}
\label{fig:10}
\end{figure}
\begin{figure}[t]
\centering
\psfrag{l-cf}[cc][][0.8]{$\mathcal{L}(f)$}%
\psfrag{l-f}[cc][][0.8]{$f$}%
\includegraphics [width=3.55in]{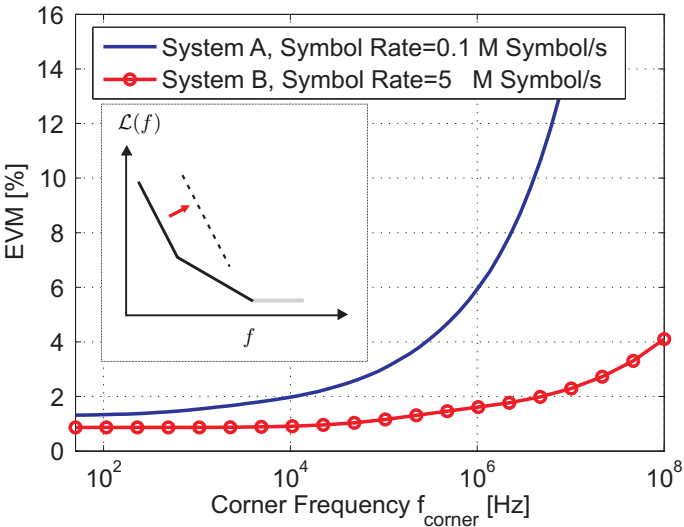}
\caption{The proposed theoretical EVM bound against different values of corner frequency $f_\tr{corner}$ for two systems with different bandwidth. $K_2=0.1$ and $K_0=-160~\tr{dBc/Hz}$ are kept constant and $f_\tr{corner}$ is increased by adding to $K_3$. The low cut-off frequency $\gamma$ is considered to be $1~\tr{Hz}$ and SNR$=30$~dB and block-length is set to $10$.}
\label{fig:9}
\end{figure}
\begin{figure}[t]
\centering
\psfrag{l-cf}[cc][][0.8]{$\mathcal{L}(f)$}%
\psfrag{l-f}[cc][][0.8]{$f$}%
\includegraphics [height=2.9in]{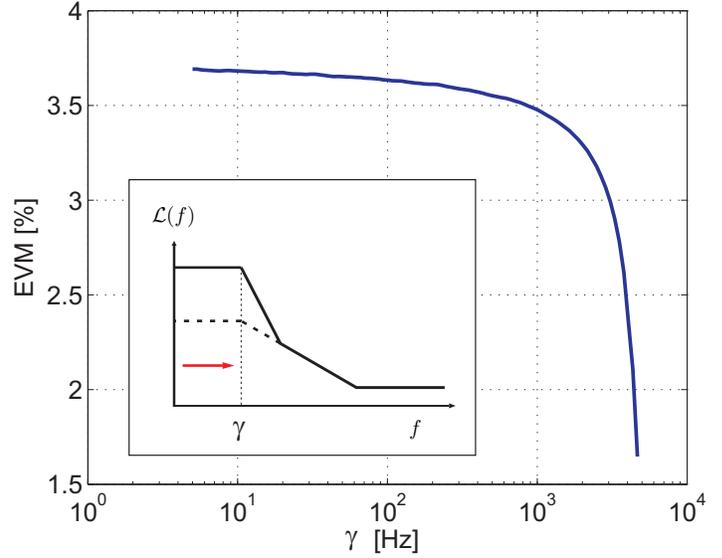}
\caption{The proposed theoretical EVM bound against different values of low cut-off frequency $\gamma$. $K_2=10^4$, $K_2=1$ and $K_0=-160~\tr{dBc/Hz}$ are kept constant, SNR$=30$~dB and symbol rate is $1$ M Symbol/s.}
\label{fig:EVMvsPLL}
\end{figure}
%Now that the EVM bound is evaluated, it can be used in order to predict the system performance for given measurement's parameters. Here, in the first subsection, effect of changing parameters in $\mathcal{L}(f)$ such as frequency figures of merit, corner frequency of $1/f^2$ and $1/f^3$, and the noise floor level on the system performance are studied. Then, these parameters are extracted from some real measurements of oscillators to compute the EVM and evaluate which oscillator can perform better in a communication system.
Now, when the EVM bound is evaluated, the system performance for a given oscillator spectrum may be quantified. In this section we study how the EVM is affected by white PN (PN floor) and cumulative PN, respectively. The effect from cumulative PN is further divided into origins from white and colored noise sources, i.e., SSB PN slopes of $-30~\tr{dB}/\tr{decade}$ and $-20~\tr{dB}/\tr{decade}$, respectively. It is found that the influence from the different noise regions strongly depends on the communication bandwidth, i.e., the symbol rate. For high symbol rates, white PN is more important compared to the cumulative PN that appears near carrier.

\begin{figure}[t]
\vspace{0.5cm}
\centering
\psfrag{Lk2}[cc][][1]{{\color[rgb]{0,0,1}$\frac{0.06}{f^2}$}}%
\psfrag{Lk3}[cc][][1]{{\color[rgb]{1,0,0}$\frac{8\times 10^3}{f^3}$}}%
\includegraphics [width=3.6in,height=2.65in]{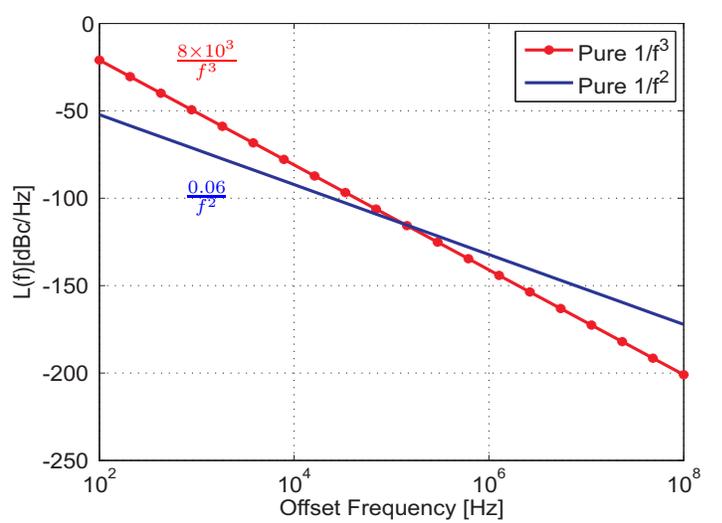}
\caption{Two SSB PN spectrums with pure $1/f^2$ PN and $1/f^3$ PN. We assume the two spectrums have a very low white PN level. For a system with the symbol rate of $3.84$~MSymbols/s, both spectrums result in the same variance of phase increments $R_\zeta(\tau=0)= 6.2\times 10^{-7}~[\tr{rad}^2]$.}
\label{fig:14}
\end{figure}

\begin{figure}[t]
\centering
\psfrag{l-f0}[cc][][0.8]{$f_0=9.85$~GHz}%
\psfrag{l-p}[cc][][0.8]{$P_\tr{osc}=-14.83$~dBc/Hz}%
\psfrag{l-k0}[cc][][0.8]{$-147.67$~dBc/Hz}%
\psfrag{l-k2}[cc][][0.9]{$\frac{0.06}{f^2}$}%
\psfrag{l-k3}[cc][][0.9]{$\frac{5\times 10^3}{f^3}$}%
\includegraphics [width=3.5in]{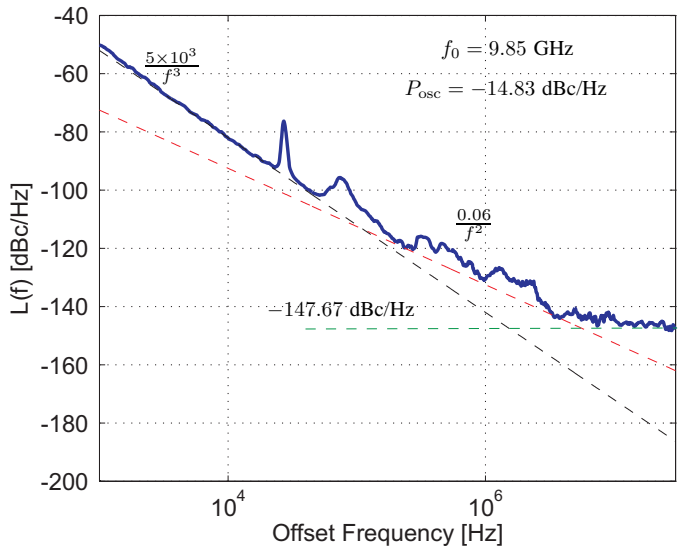}
\caption{SSB PN spectrum from a GaN HEMT MMIC oscillator. Drain voltage  $\tr{Vdd}= 6~\tr{V}$ and drain current $\tr{Id}= 30~\tr{mA}$. The corner frequency at $f_\tr{corner}=83.3$~kHz.}
\label{fig:11}
\end{figure}

\begin{figure}[t]
\centering
\psfrag{l-f0}[cc][][0.8]{$f_0=9.97$~GHz}%
\psfrag{l-p}[cc][][0.8]{$P_\tr{osc}=-7.54$~dBc/Hz}%
\psfrag{l-k0}[cc][][0.8]{$-153.83$~dBc/Hz}%
\psfrag{l-k3}[cc][][0.9]{$\frac{42\times 10^3}{f^3}$}%
\includegraphics [width=3.5in]{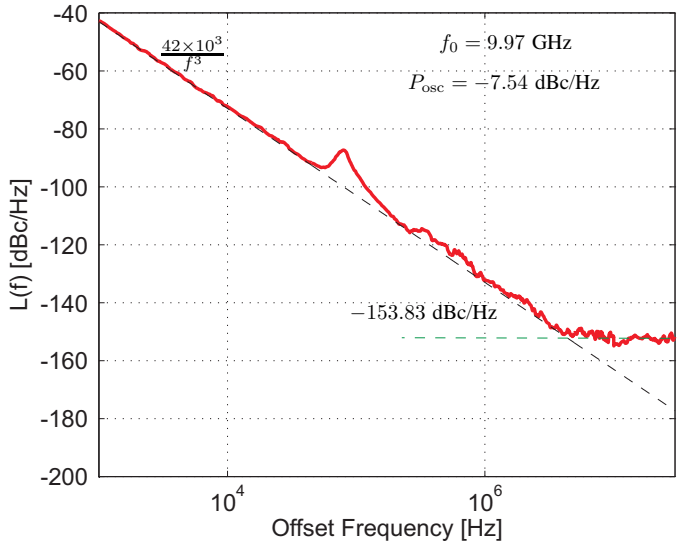}
\caption{SSB PN spectrum from a GaN HEMT MMIC oscillator. Drain voltage  $\tr{Vdd}= 30~\tr{V}$ and drain current $\tr{Id}= 180~\tr{mA}$.}
\label{fig:12}
\end{figure}

\begin{figure}[t]
\centering
\psfrag{lf1}[cc][][0.95]{\ref{fig:11}}%
\psfrag{lf2}[cc][][0.95]{\ref{fig:12}}%
\psfrag{L(f)}[cc][][0.9]{$\mathcal{L}(f)$}%
\includegraphics [width=3.35in,height=2.75in]{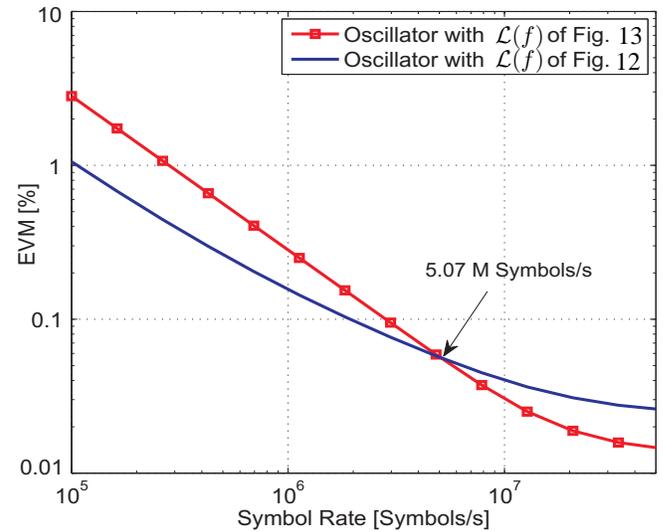}
\caption{EVM comparison of given measurements in Fig.~\ref{fig:11} and \ref{fig:12} vs. symbol rate (bandwidth). The low cut-off frequency $\gamma$  is considered to be $1~\tr{Hz}$, and SNR=$30$ dB.}
\label{fig:13}
\end{figure}
Fig.~\ref{fig:10} compares the performance sensitivity of two communication systems with different bandwidths, namely System~A and System~B against a set of different noise floor levels. System~A operates with the symbol rate of $0.1$~MSymbols/s that leads to $10~\mu$s symbol duration. In contrast, System~B has $5$~MSymbols/s symbol rate results in $0.2~\mu$s symbol time that is almost $50$ times shorter than that of System~A. It is seen in Fig.~\ref{fig:10} that an increase in the level of white PN affects the System~B  with high symbol rate much more than the more narrowband System~A system. This result can be intuitively understood, since in a system with a higher symbol rate, symbols are transmitted over a shorter period of time and thus experience smaller amount of cumulative PN. On the other hand, the amount of phase perturbation introduced by the white PN is a function of the system bandwidth and a wideband system integrates a larger amount of white PN (\ref{WhitePN_Var}). Therefore, in contrast to the cumulative PN, white PN affects a system with high bandwidth more compared to a system with a narrower bandwidth. 

The next step is to identify the different effects from cumulative PN originating in white noise sources (slope $-20~\tr{dB}/\tr{decade}$) and cumulative PN originating in colored noise sources (slope $-30~\tr{dB}/\tr{decade}$). Fig.~\ref{fig:9} shows the effect of changing the corner frequency on the performance of the introduced systems by increasing the level of $1/f^3$ noise, $K_3$. Other parameters such as $K_2$ and $K_0$ are kept constant in this simulation to just capture the effect of different values of $K_3$. Intuitively the performance degrades when the noise level is increased. However, as seen in Fig.~\ref{fig:9}, the EVM is not significantly affected below certain corner frequencies ($f_\tr{corner}<10$~kHz for System~A and $f_\tr{corner}<1$~MHz for System~B). This constant EVM is due to the dominant effect of $1/f^2$ on the performance. By increasing the corner frequency, after a certain point $1/f^3$ becomes more dominant which results in a continuous increase in EVM. It can also be seen that System~A is more sensitive to increase of the $1/f^3$ noise level. Because of the higher bandwidth, System~B contains more of the $1/f^2$ noise which is constant and dominates the $1/f^3$ effect, and its EVM stays unchanged for a larger range of corner frequencies.

Fig.~\ref{fig:EVMvsPLL} illustrates the effect of increasing the low cut-off frequency $\gamma$ on the EVM bound. As mentioned before, the PN spectrum after a PLL can be modeled similar to a free running oscillator with a flat region below a certain frequency. In our analysis, $\gamma$ is the low cut-off frequency below which the spectrum flatten. It can be seen that changes of $\gamma$ below certain frequencies ($\gamma < 1$~kHz) does not have any significant effect on the calculated EVM. However, by increasing $\gamma$ more, the effect of the flat region becomes significant and the final EVM decreases. 

Finally, we compare the individual effect of $1/f^2$ PN and $1/f^3$ PN on the performance. Consider two SSB PN spectrums as illustrated in Fig.~\ref{fig:14}. One of the spectrums contains pure $1/f^2$ PN while $1/f^3$ PN is dominant in another. In a system with the symbol rate of $3.84$~MSymbols/s (bandwidth of $3.84$~MHz), the variance of phase increment process for both spectrums is equal to  $R_\zeta(\tau=0)= 6.2\times 10^{-7}~[\tr{rad}^2]$. However, comparing the EVM values shows that the spectrum with pure $1/f^3$ PN results in $2.36$ dB lower EVM. This is due to the correlated samples of phase increment process for $1/f^3$ noise which results in lower PN estimation errors compared to $1/f^2$ noise.
\subsection{Measurements}
\label{SS_Measurements}
To materialize the analytical discussion above, Fig.~\ref{fig:11} and Fig.~\ref{fig:12} show the measured SSB PN spectrums from a GaN HEMT MMIC oscillator under two different bias conditions with drastically different characteristics for the cumulative PN. Fig.~\ref{fig:11} shows the spectrum for the oscillator biased at a drain voltage of $\tr{Vdd}= 6~\tr{V}$ and drain current of $\tr{Id}= 30~\tr{mA}$. At this bias condition, the $1/f$ noise (flicker noise) from the transistor is fairly low. The corner frequency between the $-30~\tr{dB}/\tr{decade}$ and $-20~\tr{dB}/\tr{decade}$ regions can be clearly detected at $83.3$~kHz. In contrast, Fig.~\ref{fig:12} shows a spectrum from the same oscillator biased at $\tr{Vdd}= 30~\tr{V}$ and $\tr{Id}= 180~\tr{mA}$. Under this bias condition the noise from colored noise sources is increased significantly and the cumulative PN has  $-30~\tr{dB}/\tr{decade}$  slope until it reaches the white PN floor. Further, the power of the oscillator is higher in Fig~\ref{fig:11}, resulting in a lower level for the white PN. Fig.~\ref{fig:13} compares EVM for the SSB PN spectrums in Figs.~\ref{fig:11} and \ref{fig:12}, respectively, versus the symbol rate. As expected based on the results in Sec.~\ref{SS_Discussions}, the spectrum in Fig~\ref{fig:11} gives the best EVM for low symbol rates, while the spectrum in Fig~\ref{fig:12} gives the best EVM for higher symbol rates as a result of the lower level of white PN. 
%Using the results of Sec.~\ref{Sec_PhaseNoiseModel}, at the symbol rate of $5.07~\tr{MSymbols/s}$ (bandwidth of $5.07~\tr{MHz}$) the variance of JAP for the spectrums in Fig~\ref{fig:11} and \ref{fig:12}  is equal to $ R_\zeta(\tau=0)= 6.9\times 10^{-7}~[\tr{rad}^2]$ and $ R_\zeta(\tau=0)= 1.9\times 10^{-6}[\tr{rad}^2]$ ,  respectively. Although the oscillator with spectrum of Fig.~\ref{fig:12} has a higher JAP variance and also a higher white noise floor level, it has similar EVM to the other oscillator. This is due the dominance of $1/f^3$ noise in this spectrum with correlated JAP samples (see (\ref{f3_ACF_1})). Exploiting this correlation results in lower residual PN estimation errors compared to $1/f^2$ noise that is originated from uncorrelated JAP samples.
\section{Conclusions}
\label{Sec_Conclusions}
In this paper, a direct connection between oscillator measurements, in terms of measured single-side band PN spectrum, and the optimal communication system performance, in terms of EVM, is mathematically derived and analyzed. First, we found the statistical model of the PN which considers the effect of white and colored noise sources. Then, we utilized this model to derive the modified Bayesian Cram\'{e}r-Rao bound on PN estimation that is used to find an EVM bound for the system performance.

The paper demonstrates that for high symbol rate communication systems, the near carrier cumulative PN is of relatively low importance compared to white PN far from carrier. Our results also show that $1/f^3$ noise is more predictable compared to $1/f^2$ noise, and in a fair comparison it affects the system performance less. These findings will have important effects on design of hardware for frequency generation as well as the requirements on voltage controlled oscillator design, choice of reference oscillators and loop bandwidth in the phase-locked loops.

Although in several empirical measurements of oscillators $1/f^3$, $1/f^2$, and $f^0$-shaped noise dominate the PN spectrum, there has been studies where other slopes ($1/f^4$, $1/f^1$) have been observed in the measurements. Our PN model can be extended in future studies to include the effect of various noise statistics. Our current analysis can be used in order to study free running oscillators, and it is valid for study of phase-locked loops up to some extent. Further, our theoretical results can be extended for a more thorough study of  phase-locked loops. In our analysis, the transition from continuous to discrete-time domain was based on a slow-varying PN assumption.  A more sophisticated study can be conducted to analyze the effect of relaxing this assumption. Finally, we analyzed a single carrier communication system. It is interesting to extend this work to the case of multi-carrier communication systems.

\begin{appendices}
\section{}
\label{Appendix_B}
In this appendix, elements of the covariance matrix $\mathbf{C}$ are calculated. According to (\ref{phase_increament_dis2}), and stationarity of the PN increments  
\begin{align}
\label{CovMatrix_appendix_0}
%\phi_\tr{c}(1)+\sum_{i=1}^{n-1}\zeta(i)+\phi_0[n]\\
[\mathbf{C}]_{l,k}&=\mathbb{E}\Big[\big(\phi[l]-\mathbb{E}[\phi[l]]\big)\big(\phi[k]-\mathbb{E}[\phi[k]]\big)\Big]\nonumber\\
&=\mathbb{E}\Big[\Big(\phi_0[l]+\phi_3[1]+\phi_2[1]+\sum_{m=2}^{l}(\zeta_3[m]+\zeta_2[m])\Big)\nonumber\\
&\hspace{0.9cm}\times\Big(\phi_0[k]+\phi_3[1]+\phi_2[1]+\sum_{m'=2}^{k}(\zeta_3[m']+\zeta_2[m'])\Big)\Big]\nonumber\\
&=\mathbb{E}\Big[\phi_3[1]\phi_3[1]\Big]+\mathbb{E}\Big[\phi_2[1]\phi_2[1]\Big]+\mathbb{E}\Big[\phi_0[l]\phi_0[k]\Big]\nonumber\\
&\hspace{0.9cm}+\sum_{m=2}^{l}\sum_{m'=2}^{k}\mathbb{E}\Big[\zeta_3[m]\zeta_3[m']\Big]+\mathbb{E}\Big[\zeta_2[m]\zeta_2[m']\Big]\nonumber\\
&\hspace{0.9cm}+\mathbb{E}\Big[\phi_3[1]\times\sum_{m=2}^{l}\zeta_3[m]\Big]+\mathbb{E}\Big[\phi_3[1]\times\sum_{m'=2}^{k}\zeta_3[m']\Big]\nonumber\\
&\hspace{0.9cm}+\mathbb{E}\Big[\phi_2[1]\times\sum_{m=2}^{l}\zeta_2[m]\Big]+\mathbb{E}\Big[\phi_2[1]\times\sum_{m'=2}^{k}\zeta_2[m']\Big],\nonumber\\
&l,k=\{1\dots N\}.
\end{align}
Note that in calculation of the MBCRB, we need to compute the inverse of the covariance matrix $\mathbf{C}$. It is possible to mathematically show that the correlations between the initial PN of the block and future PN increments (the four last terms in (\ref{CovMatrix_appendix_0})) do not have any effect on $\mathbf{C}^{-1}$. Therefore, we omit those terms in our calculations and finally the covariance matrix can be written as
\begin{align}
\label{CovMatrix_appendix}
\hspace{-0.8cm}[\mathbf{C}]_{l,k}&=\sigma^2_{\phi_3[1]}+\sigma^2_{\phi_2[1]}+\delta[l-k]\sigma_{\phi_0}^2\nonumber\\
&\hspace{0.8cm}+\sum_{m=2}^{l}\sum_{m'=2}^{k} R_{\zeta_3}[m-m']+R_{\zeta_2}[m-m'],\nonumber\\
&l,k=\{1\dots N\}.
\end{align}
\section{}
\label{Appendix_A}
Steps taken to solve the integral in (\ref{f3_PSD_Cont}) are described here. We can write (\ref{f3_PSD_Cont}) as
\begin{align}
\label{integral_1}
R_{\zeta_{3}}(\tau)&=8\int^{+\infty}_{0}\frac{K_3}{f^3+\gamma^3}\sin(\pi fT)^2 \cos( 2\pi f \tau)\tr{d}f\nonumber\\
&=4\int^{+\infty}_{0}\frac{K_3}{f^3+\gamma^3}\cos( 2\pi f \tau)\tr{d}f\nonumber\\
&\quad-2\int^{+\infty}_{0}\frac{K_3}{f^3+\gamma^3}\cos( 2\pi f (|\tau +T|)\tr{d}f\nonumber\\
&\quad-2\int^{+\infty}_{0}\frac{K_3}{f^3+\gamma^3}\cos( 2\pi f (|\tau -T|)\tr{d}f.
\end{align}
It is clear that solving the integral in the form of $\int^{+\infty}_{0}{K_3}/{(f^3+\gamma^3)}\cos( 2\pi f \tau)\tr{d}f$ is enough to compute the total integral of (\ref{integral_1}). This integral is complicated enough that powerful software such as Mathematica are not able to converge to the final answer. Consequently, first, partial-fraction decomposition of ${1}/{(f^3+\gamma^3)}$ is done: 
\begin{align}
\label{frac_decom_1}
&\frac{1}{f^3+\gamma^3}=\frac{A}{f-(-\gamma)}+\frac{B}{f-\gamma e^{j\pi/3}}+\frac{C}{f-\gamma e^{-j\pi/3}}\nonumber\\
&A=\frac{1}{3\gamma^2},~B=\frac{e^{-j2\pi/3}}{3\gamma^2},~C=\frac{e^{j2\pi/3}}{3\gamma^2}
\end{align}
Note that, $\gamma$ is a real positive number. Using Mathematica (Version 7.0), the following integral can be evaluated
\begin{align}
\label{general_int}
\int^{+\infty}_{0}\frac{1}{f+\beta}\cos( 2\pi f \tau)\tr{d}f=\nonumber\\
&\hspace{-2.5cm}-\cos(2\beta\pi\tau)\tr{cosint}(-2\beta\pi|\tau|)\nonumber\\
&\hspace{-2.5cm}-\frac{1}{2}\sin(2\beta\pi|\tau|)(\pi+2~\tr{sinint}(2\beta\pi|\tau|)),
\end{align}
where $\beta$ must be a complex or a negative real number, and $\tr{sinint}(\cdot)$ and $\tr{cosint}(\cdot)$ are sine and cosine integrals define as
\begin{align}
\label{cos_sin_int}
&\tr{sinint}(x)=\int^{r}_{0}\frac{\sin(t)}{t}\tr{d}t\nonumber\\
&\tr{cosint}(x)=\Gamma+\log(x)+\int^{x}_{0}\frac{\cos(t)-1}{t}\tr{d}t,
\end{align}
where $\Gamma\approx 0.5772$ is the Euler-Mascheroni's constant. 

Consider the case where time lag $\tau$ is small. By Taylor expansion of the functions in (\ref{general_int}) around zero
\begin{align}
&\sin(x)=x-\frac{x^3}{6}+\dots\nonumber 
&&\cos(x)=1-\frac{x^2}{2}+\dots\\
&\tr{sinint}(x)=x - \frac{x^3}{18} + \dots\nonumber
&&\tr{cosint}(x)=\Gamma+\log(x)-\frac{x^2}{4}+\dots,
\end{align}
and neglecting the terms after second order, the integral can be approximated as
\begin{align}
\label{approx_int}
\int^{+\infty}_{0}\frac{1}{f+\beta}\cos( 2\pi f \tau)\tr{d}f\approx\nonumber\\
&\hspace{-2cm}-\Gamma-\log(-2\beta\pi|\tau|)-2\beta\pi^2|\tau|\nonumber\\
&\hspace{-2cm}+\frac{(2\beta\pi|\tau|)^2}{2}(\Lambda+\log(-2\beta\pi|\tau|)),
\end{align}
where $\Lambda\triangleq\Gamma-\frac{3}{2}$.
Employing this approximation and the fraction decomposition in (\ref{frac_decom_1}), followed by a series of simplifications
\begin{align}
\int^{+\infty}_{0}\frac{K_3}{f^3+\gamma^3}\cos( 2\pi f \tau)\tr{d}f\approx\nonumber\\
&\hspace{-3cm}\frac{K_3}{3\gamma^2}\left(\frac{2\pi}{\sqrt{3}}+6\gamma ^2\pi^2\tau^2(\Lambda+\log(2\gamma \pi|\tau|))\right). 
\end{align}
Now the first term in (\ref{integral_1}) is calculated. By changing the variable $\tau$ to $\tau+T$ and $\tau-T$, second and third terms can also be computed, respectively. Finally, $R_{\zeta_{3}}$ is approximated by
\begin{align}
\label{final_ACF_f3_app}
R_{\zeta_{3}}(\tau)\approx&-8K_3\pi^2 \Big[-\tau^2(\Lambda+\log(2\pi\gamma |\tau|))\nonumber\\
&+\frac{(\tau+T)^2}{2}(\Lambda+\log(2\pi\gamma |\tau+T|))\nonumber\\
&+\frac{(\tau-T)^2}{2}(\Lambda+\log(2\pi\gamma |\tau-T|))\Big].
\end{align}
To calculate the ACF for $\tau=0$, and $\tau=|T|$, we need to take the limits of (\ref{final_ACF_f3_app}) as $\tau$ approaches $0$, and $|T|$, respectively that results in
\begin{align}
\label{lim_0_final_ACF_f3_app}
&\hspace{-0.5cm}\lim_{\tau \to 0} R_{\zeta_3}(\tau)\approx-8K_3\pi^2 T^2(\Lambda+\log(2\pi\gamma T)),
\end{align}
\begin{align}
\label{lim_dT_final_ACF_f3_app}
&\hspace{-0.5cm}\lim_{\tau \to |T|} R_{\zeta_3}(\tau)\approx-8K_3\pi^2 T^2(\Lambda+\log(8\pi\gamma T)).
\end{align}
\end{appendices}
\bibliographystyle{IEEEtran}
%\bibliography{IEEEabrv,references}

\end{document}